\documentclass[prc,aps,nofootinbib,showkeys,showpacs,twocolumn,floatfix]{revtex4}   
\usepackage{epsfig}  
\usepackage{graphicx} 
\usepackage{amsmath}
\usepackage{amssymb}
\usepackage{amsfonts}
\usepackage{color}
\usepackage{color}  
\graphicspath{{../}} 
\begin{document}  

\date{\today}
\title{
Deformation effects on the coexistence between neutron-proton and particle like pairing in N=Z medium mass nuclei
}

\author{Danilo Gambacurta}\email{danilo.gambacurta@ct.infn.it}
\affiliation{Istituto Nazionale di Fisica Nucleare, Sezione di Catania, Via S. Sofia 64, I-95123 Catania, Italy}
\affiliation{GANIL, CEA/DSM and CNRS/IN2P3, Bo\^ite Postale 55027, 14076 Caen Cedex, France}  
\author{Denis Lacroix} \email{lacroix@ipno.in2p3.fr}  
\affiliation{Institut de Physique Nucl\'eaire, IN2P3-CNRS, Universit\'e Paris-Sud, F-91406 Orsay Cedex, France}   

\begin{abstract}
A model combining self-consistent mean-field and shell-model techniques is used to study the competition between particle like and proton-neutron  pairing correlations in fp-shell even-even self-conjugate nuclei. Results obtained using constant two-body pairing interactions as well as more sophisticated interactions  are presented and discussed. The standard BCS calculations are
systematically compared with more refined approaches including correlation effects beyond the independent quasi-particle approach. The competition between proton-neutron correlations in the isoscalar and isovector channels
is also analyzed, as well as their dependence on the deformation properties. 
Besides the expected role of the spin-orbit interaction and particle number conservation,  
it is  shown that deformation leads to a reduction of the pairing correlations. 
This reduction originates from the change of the single-particle spectrum and from a quenching of the residual pairing matrix elements. 
The competition between isoscalar and isovector pairing in the deuteron transfer is finally addressed.
Although a strong dependence the isovector pairing correlations with respect to nuclear deformation is observed, they always dominate over the isoscalar ones.
\end{abstract}

\date{\today}



\maketitle

\section{Introduction}

Although the role of pairing correlations in nuclei was introduced 
more than 60 years by  Bohr, Mottelson and Pines \cite{Boh58} and they 
dictate many nuclear properties  \cite{Rin80,Bri05}, some aspects of pairing remain unclear. 
For instance, the precise role of the neutron (n)-proton (p) pairing in nuclei still 
challenges both theoretical and experimental nuclear physics \cite{Goo72,Goo99,Mac00,Mac00a,Lis03,Ced11,Afa2013} (see
 Ref. \cite{Fra14} for a recent review). The effect of  correlation between nucleons of different spin and isospin, is expected to be more pronounced in self-conjugate nuclei. From an experimental point of view,  high intensity radioactive beams will offer new possibilities to study the importance of isoscalar (T = 0) and isovector (T = 1) pairing interaction between protons and neutrons along the
N = Z line. The role of isovector proton-neutron (p-n) pairing correlations has been recently pointed out
by analyzing the relative energies of the T=0 and T=1 states in even-even and
odd-odd nuclei \cite{Macc2000} and the T = 0 band in $^{74}$Rb \citep{74Rb}. 
The analysis based on these results provides evidences of the existence of a neutron-proton isovector pair field but does not support the existence of the isoscalar one.
Conjointly, recent experiments seem to manifest the possibility to observe exotic structure  of aligned pairs \cite{Ced11} that could be explained in terms of isoscalar p-n  pairing 
correlations. 
 \begin{figure*}[htb]
\begin{center}
\includegraphics[width = 16.cm]{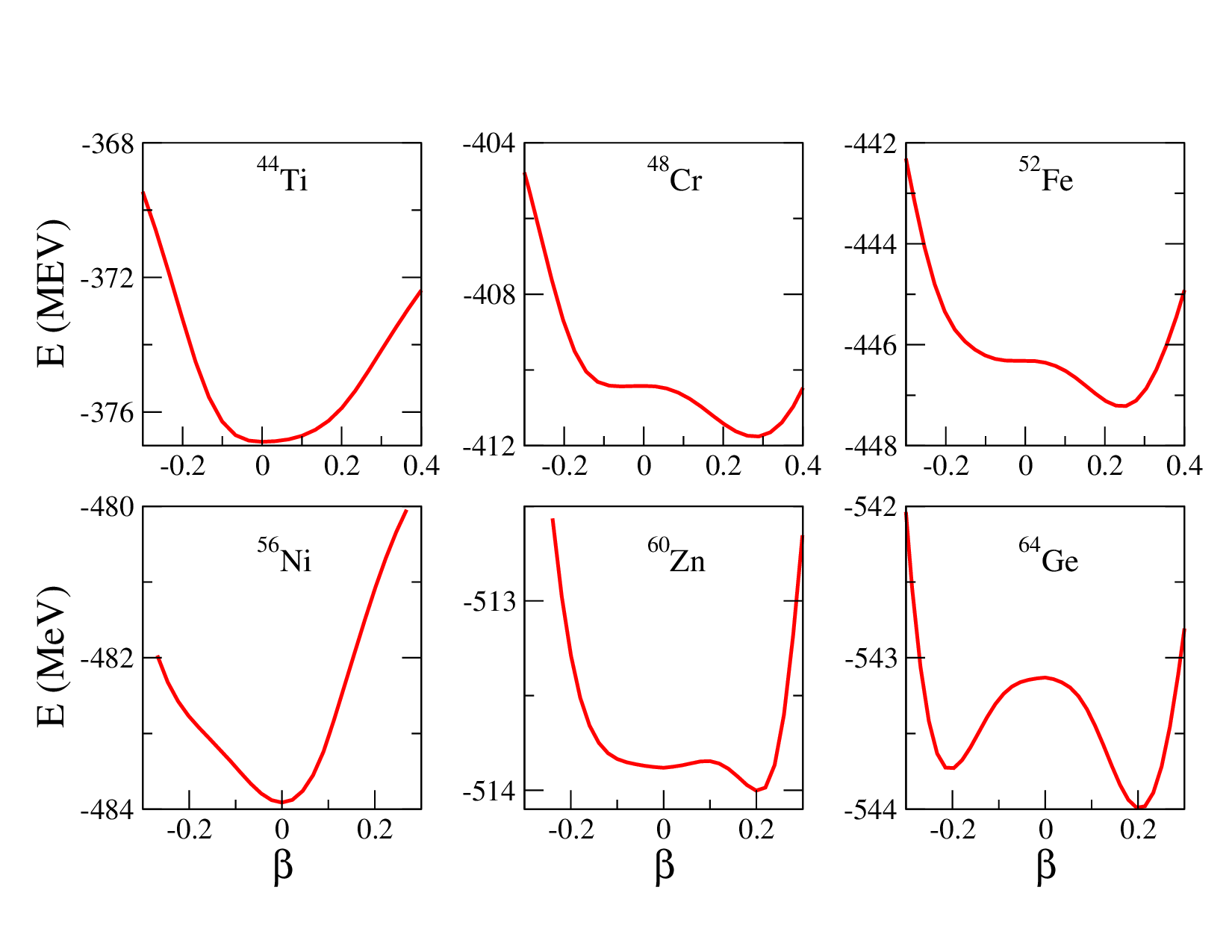}
\end{center}
\caption{(Color online) Binding energy as a function of the
quadrupole deformation parameter obtained in the HF+BCS calculations 
for different $Z=N$ nuclei (see text for details).}
\label{Fig:EvsBeta}
\end{figure*}  

 From a theoretical point of view, several frameworks have been proposed to incorporate p-n pairing  correlations in microscopic models. 
 Many works devoted to the study of the competition between isoscalar and isovector pairing,  
 have been performed in solvable models (see for example \citep{Par65,Evans1981,Dussel1986,Engel1997,Dobes1998,Lerma2007}).  Mean-field approaches, generally only incorporate particle-like pairing correlations. Extensions to include p-n correlations have been already proposed some times ago \cite{Goo72}
 and sometimes applied  in the Hartree-Fock Bogolyubov (HFB) approach \cite{Ter98,Ber10}. Most often, these approaches lead to a non-coexistence of particle-like and particle unlike pairing that is further supported by the analytical work in 
Refs. \cite{Gin68,San09}. It is worth mentioning that such coexistence has been found away from stability 
in some exotic situations \cite{Gez11}. Alternatively, shell-model calculations starting from a simplified pairing Hamiltonian 
can go beyond the independent quasi-particle picture and provide a particle number conserving framework able to attack the pairing problem 
including all spin/isospin channels. Beyond mean-field studies have been recently performed  to study the competition between $T=0$ and $T=1$ 
pairing in spherical nuclei \cite{Sag13}, to understand the origin of the Wigner energy \cite{Ben13,Ben14} to probe the existence of quarteting \cite{San12a,San12b,San2014,Samb2014},
to describe spin-aligned pairs \cite{Isacker2011} or deuteron transfer properties in N=Z nuclei
\cite{Isacker2005}.

It is worth mentioning that deformation has sometimes been included using schematic \cite{Lei11} or more realistic Hamiltonian \cite{Leb12}.
The aim of the present work is tomake a precise study of the role of deformation on particle-like and p-n pairing by using the following strategy. A microscopic mean-field is used to obtain realistic single-particle energies and two-body residual pairing interactions. Then,  pairing correlations 
are studied in spherical and deformed nuclei through direct diagonalization of the Hamiltonian in a restricted space.  This framework, that combines the self-consistent mean-field approach and the full diagonalization of the pairing Hamiltonian, allows for a realistic description of  nuclear deformation necessary to analyze its influence on the role of pairing correlations in both channels. 
\begin{figure*}[htb]
\begin{center}
\includegraphics[width = 16.cm]{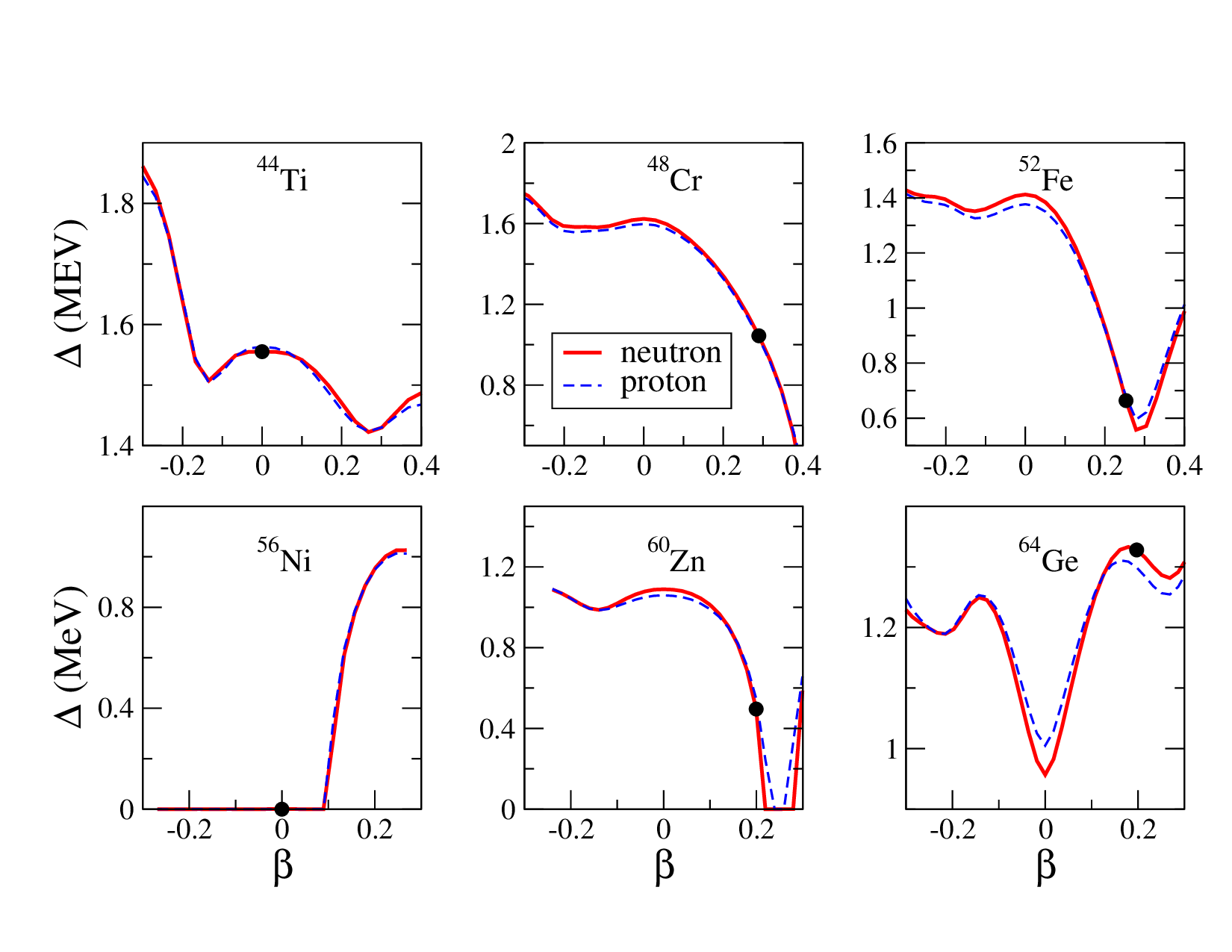}
\end{center}
\caption{(Color online)  Mean proton (blue dashed line) and neutron (red full line) gaps as a function of the average quadrupole deformation parameter $\beta$.
For each nucleus, the black solid circle indicates the equilibrium configuration that minimizes the EDF.
}
\label{Fig:GapvsBeta}
\end{figure*} 

\begin{figure}[htbp]
\begin{center}
\includegraphics[width = 8.cm]{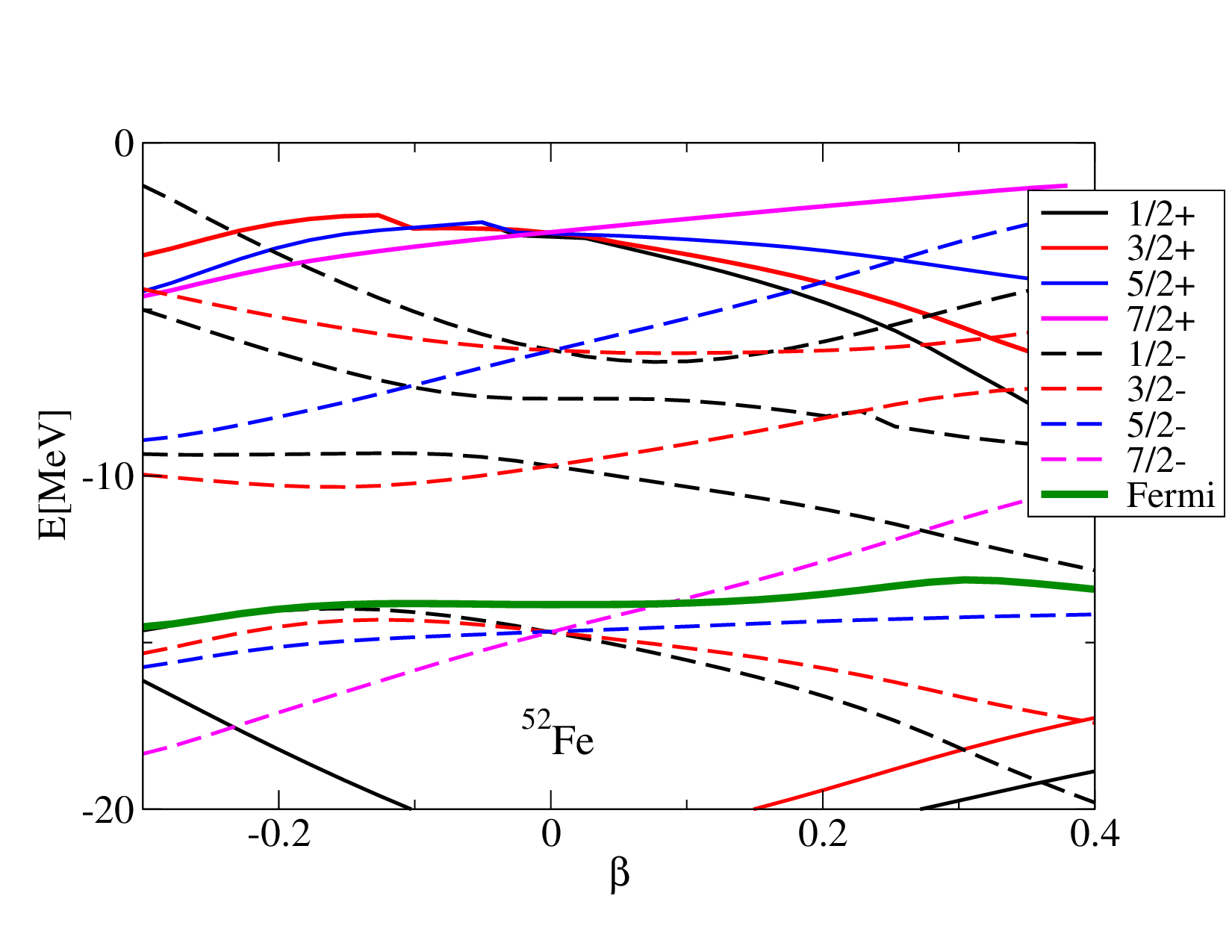}
\end{center}
\caption{(Color online) Evolution of the neutron single-particle energies as a function of the deformation $\beta$ obtained in the mean-field
calculations for $^{52}$Fe. The calculation are performed using {\sl EV8} with only particle-like pairing. 
Negative and positive parity states are plotted with solid and dashed lines, respectively. The thick solid green
line indicates the Fermi energy. Note that here, the single-particle quantum numbers have been assigned by continuity with the spherical case.}
\label{Fig:Esp-Fe52}
\end{figure} 

\begin{figure}[htbp]
\begin{center}
\includegraphics[width = 8.cm]{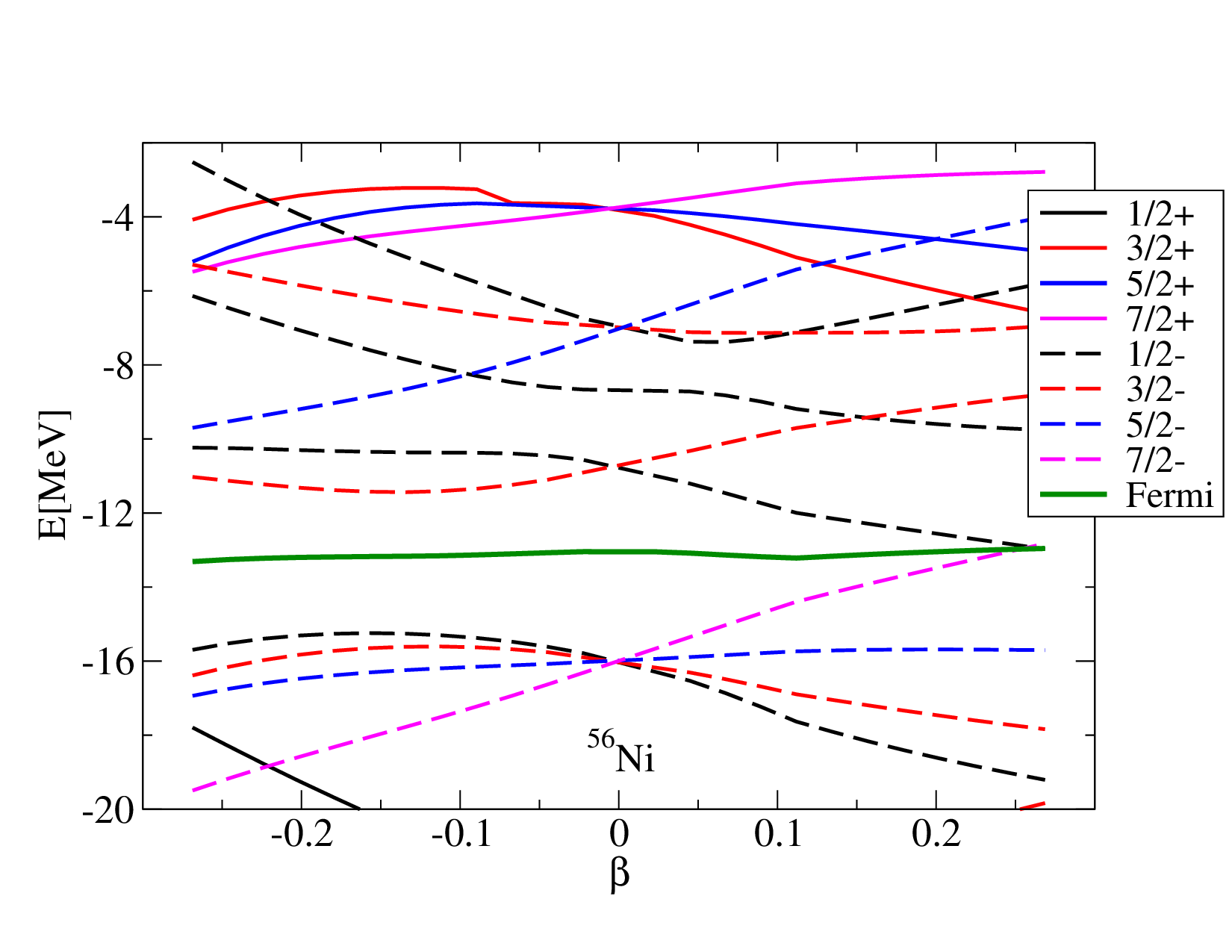}
\end{center}
\caption{(Color online) Same as Fig. \ref{Fig:Esp-Fe52} for $^{56}$Ni}
\label{Fig:Esp-Ni56}
\end{figure} 
The paper is organized as follows. In Section \ref{Sec:mf}, the results obtained in the Skyrme-Hartre-Fock (HF)+BCS mean-field calculations for different fp-shell $N=Z$  nuclei are presented and discussed in terms of the total binding energy, mean pairing gap and single-particle energies dependence on the deformation. In  
Section \ref{Sec:sm}, the methodology used to connect the mean-field and the shell-model (SM) techniques  is discussed. In  Section \ref{Sec:Gconst}, some results obtained with a constant residual interaction in the isovector channel will be shown. In Section \ref{Sec:Greal},
more realistic interactions are used in the two channels to study
the interplay between deformation and pairing in various channels.   Finally, in  Section \ref{Sec:concl} some conclusions will be drawn.

 \section{Preliminary study: Mean-field description of medium mass N=Z nuclei}\label{Sec:mf}

We will focus here on fp-shell $N=Z$ even-even nuclei. To illustrate the importance of deformation 
in this region of mass, we have systematically applied the  {\sl EV8} code \cite{ev8}.
The self-consistent mean-field equations are solved in a three-dimensional mesh using the zero-range Skyrme Energy density functional (EDF) in the mean-field 
channel. In addition, pairing correlations between neutrons  or protons  are accounted for using the Hartree-Fock +BCS approximation. In the present work, the SLy4 parametrization \cite{sly4} 
is used in the particle-hole channel while a density-dependent residual contact
interaction
\begin{equation}
 V(\textbf{r},\textbf{r'})=-v_0 \left(1-\eta\frac{\rho(\textbf{r})}{\rho_0} \right)\delta(\textbf{r}-\textbf{r'}),
 \label{eqn:vpair}
\end{equation} 
is used in the pairing channel. Results shown below have been obtained using a mixed-type interaction ($\eta=0.5$, $v_0=700$ MeV fm$^{3}$, $\rho_0$=0.16 fm$^{-3}$) taken from \cite{Alh2011} that has been adjusted to
reproduce pairing gaps along the nuclear chart.

An illustration of the total energy evolution for different even-even $N=Z$ nuclei, with  $N=$22 to $N=$32, as a function of the quadrupole  
deformation parameter $\beta$,  is given in  Fig. \ref{Fig:EvsBeta}.  The deformation parameter is defined as  
\begin{equation}
 \beta= \left(\frac{5\pi}{9} \right)^\frac{1}{2} \frac{\langle \hat Q_2 \rangle}{A R_0^2},
\label{Eq:beta}
\end{equation} 
where  $A$ is the mass number, $R_0=1.2 A^\frac{1}{3}$ and $\hat Q_2$ the quadrupole operator.
Except the $^{44}$Ti and $^{56}$Ni that are found to be spherical, all considered nuclei are predicted prolate 
with deformation parameters around 0.2 and more or less pronounced minima in the potential energy landscape.
\subsection{Evolution of particle-like pairing with deformation}

Already at the mean-field level, significant deformation effects on the pairing correlation strength are seen.
Using standard notations \cite{Rin80}, we denote by $(u_i,v_i)$ the BCS coefficients. In Fig.  \ref{Fig:GapvsBeta} the average proton and neutron gaps 
defined as
\begin{equation}
 \Delta=\frac{\sum_i\Delta_i v_i^2 }{\sum_i v_i^2} \nonumber 
\end{equation} 
where $\Delta_i = \sum_j V^{10}_{ij}u_jv_j$ ($V^{10}_{ij}$ being the matrix element of
the pairing interaction in the $T=1,S=0$ channel, see below) is the pairing gap of the single-particle level $i$,
are plotted   as a function of  $\beta$.
It could be first noted that the proton and neutron pairing gaps are almost identical. This is due to the quasi-isospin symmetry 
in the $N=Z$ nuclei. Note that, here, the isospin symmetry is explicitly broken due to the Coulomb interaction. 

This figure also illustrates the large fluctuations of the pairing correlation as a function of deformation. This is a well known effect that 
essentially stems from the evolution of single-particle shells; and more specifically shell gaps, with deformation. In the considered nuclei, we see that there is no systematic rules; deformation could enhance or reduce pairing compared to the spherical configuration.  

This diversity in deformation effect can be directly understood by focusing on the single-particle energies evolution as a function of $\beta$.
Such evolutions are shown for $^{52}$Fe (deformed) and $^{56}$Ni (spherical) in Figs  \ref{Fig:Esp-Fe52} and \ref{Fig:Esp-Ni56}, respectively. For  $^{52}$Fe (resp. $^{56}$Ni) a
reduction (resp. an increase) of pairing correlation is seen when deformation is non-zero compared to the spherical symmetric case. Such evolution 
can be understood as follow.
The simplest situation is the $^{56}$Ni, 
where the spherical configuration is stabilized by the $N=Z=28$ shell closure associated to a completely filled $f_{7/2}$ shell. This shell closure induces 
a pronounced gap between the   $f_{7/2}$ and the next unoccupied single-particle levels. As a consequence, the pairing gap vanishes. When deformation 	increases, the gap disappears and pairing can built-up with neighboring shells.  In  $^{52}$Fe, the situation is slightly more complex. In the spherical case, the degenerate $f_{7/2}$ is partially occupied and therefore pairing is non-zero due to the interaction of particles in the same shell. The reduction of pairing 
as $\beta$ increases is essentially due to the splitting of the single-particle $f_{7/2}$ states. 

\section{Construction of a shell-model Hamiltonian from self-consistent mean-field outputs}\label{Sec:sm}
As mentioned above, standard mean-field approaches do not lead in general to a coexistence of particle-like and p-n pairing. This is mainly due to the explicit violations of the particle number and isospin symmetry. To describe p-n pairing and more specifically its coexistence with particle-like pairing, it is advantageous to perform a many-body SM calculation beyond the independent particle or quasi-particle picture. 
Starting from the mean-field calculation, we first construct a two-body pairing Hamiltonian written as:
\begin{eqnarray}
H = H_{\rm s.p.} + H_{10} + H_{01} \label{eq:hamil}
\end{eqnarray}  
where $H_{\rm s.p.}$ is the independent particle contribution. $H_{10}$ and $H_{01}$ correspond to a pairing 
two-body Hamiltonian acting respectively in the $(T=1,S=0)$ and $(T=0,S=1)$ channels.

In the present work, we use the output of the {\sl EV8} model to construct this Hamiltonian  that realistically accounts 
for deformation. For each deformation, after the minimization process of the EDF, a set of protons and neutrons single-particle 
wave-functions are obtained. The corresponding wave-function, denoted by $| k, \tau_k \rangle$ where $\tau_k$ is the isospin quantum numbers, 
are associated to the proton and neutron creation operators $\pi^{\dagger}_k$ and $\nu^{\dagger}_k$ respectively (see   appendix \ref{app:2body} and \cite{ev8}). 
Note that, in {\sl EV8} 
time-reversal symmetry is assumed. We denote by $\bar k$ the time-reversal state associated to $k$. 

In $N=Z$ nuclei, we do expect that isospin symmetry is almost respected. This is illustrated for instance 
in Fig. \ref{Fig:GapvsBeta} where the proton and neutron gaps are almost identical. For the sake of simplicity, the isospin symmetry could be explicitly enforced. This could be done for instance by neglecting the Coulomb interaction at the mean-field level. However, since the Skyrme parameterization has been adjusted with Coulomb, the total energy dependence on deformation might be unrealistic if Coulomb is completely 
removed. 
For this reason, we preferred to keep the Coulomb interaction 
and assume a posteriori that the proton wave-functions and single-particle energies $\epsilon_k$ are identical to the neutron 
ones, i.e. $\pi_k \equiv \nu_k$.   Then, each level is 4 times degenerated and the single-particle hamiltonian writes:
 \begin{eqnarray}
 H_{s.p.}&=&\sum_{k} \epsilon_{k} (\pi^\dagger_k \pi_k + \pi^\dagger_{\bar k} \pi_{\bar k}+\nu^\dagger_k \nu_k + \nu^\dagger_{\bar k} \nu_{\bar k})
 \nonumber
\label{eqn:sp}
\end{eqnarray}
where $\epsilon_{k}$ denote the single-particle energies obtained in the mean-field solution with the {\sl EV8} code.
The two-body part of the Hamiltonian can also be constructed consistently with the pairing treatment in the {\sl EV8} code. In this model, 
since only time-reversal pairs $(k , \bar k)$  of the same isospin can form a Cooper pair,  only $T=1$, $T_z = \pm 1$ are approximately treated in the BCS approximation. 
In order to account for all isospin channels in $T=1$, we consider a two-body hamiltonian $H_{10}$ that also includes p-n interaction through
\begin{eqnarray} 
H_{10} = \sum_{i\neq j,T_z} V^{10}_{ij} 
 P^\dagger_{T_z}(i) P_{T_z}(j). 
\label{Eq:H10}
\end{eqnarray}
The different operators $P_{T_z}$ with $T_z=-1$, $0$, $+1$, denote the creations  operators of $S=0$ pairs and different isospin projections. These states 
can directly be written in terms of the time-reversed states provided by  {\sl EV8} as:
\begin{eqnarray}
P^\dagger_{1}(k) &=&  \nu^\dagger_k \nu^\dagger_{\bar k}  , ~~P^\dagger_{-1}(k) =  \pi^\dagger_k \pi^\dagger_{\bar k}, \nonumber  \\
P^\dagger_{0}(k) &=& (\nu^\dagger_k \pi^\dagger_{\bar k} + \pi^\dagger_k \nu^\dagger_{\bar k})/\sqrt{2} .  \nonumber  
\end{eqnarray}
 The two-body interaction matrix elements do not depend on the specific isospin projection $T_z$ due to the imposed symmetry between protons and
neutrons. In this paper, some applications using a constant two-body interaction will first be considered (see Section \ref{Sec:Gconst}), and then realistic two-body interactions that account in particular for the influence of deformation in single-particle states will be used (see Section \ref{Sec:Greal} and 
Appendix \ref{app:2body}). In the latter case, the two-body interaction 
matrix elements in the $T=1$ channel are computed using the same residual interaction as in the mean-field level.  

In our scheme, it is possible to identify pairs of time-reversed states that enters into the $T=1$ channels as well as in the $(T=0,S_z=0)$ channel. 
Due to the specific space symmetry used in the {\sl EV8 } code, the interaction matrix elements in $T=0$, $S_z = \pm 1$ cannot be easily restricted to $J=0$ and $J=1$ channels that are expected to be the dominant channels \citep{Goodman1972}. 
For this reason, only the $S_z = 0$ component is considered here leading to  a simplified isoscalar pairing hamiltonian:
\begin{eqnarray}
H_{01} = \sum_{i\neq j} V^{01}_{ij} D^\dagger_{0}(i) D_{0}(j),
\label{Eq:H01}
\end{eqnarray}  
where the pair creation operator $D^\dagger_{0}$ is given by
\begin{eqnarray}
D^\dagger_{0}(k) &=&  (\nu^\dagger_k \pi^\dagger_{\bar k} - \pi^\dagger_k \nu^\dagger_{\bar k})/\sqrt{2}.  \nonumber
\end{eqnarray}
In order to compute the two-body interaction components $V^{01}_{ij}$, we use the same effective interaction (\ref{eqn:vpair}) except that
the coupling strength is replaced by $v_1 = x v_0$ where $x$ is a constant. Using arguments based on shell model hamiltonians, it was shown that a realistic value for $x$ is around 1.6 \cite{Ber10}. 

Different types of SM calculations will be performed in order to single out specific effects:
\begin{itemize}
  \item  particle-like pairing: in most standard mean-field 
approaches, only n-n and p-p pairing 
  are usually treated self-consistently.  This  calculation 
will be helpful for comparison with the mean-field approach and 
  see the effect beyond the BCS technique in a particle conserving approach to 
pairing. This amounts to use the Hamiltonian
\begin{eqnarray} 
H &=& H_{s.p.}+ \sum_{i\neq j,T_z=\pm1} V^{10}_{ij} 
 P^\dagger_{T_z}(i) P_{T_z}(j) 
\label{Eq:plikeH}
\end{eqnarray}

 and the corresponding SM calculations will be denoted as  $|T_z|=1$;
  \item full isovector-pairing: this calculation will allow to understand the 
interplay between particle like and p-n pairing in the isovector channel.
  Note that, we will always consider p-n symmetry in the inputs of 
the SM calculation leading to equal contributions of the three 
  components. The Hamiltonian writes
\begin{eqnarray} 
H &=& H_{s.p.}+ \sum_{i\neq j,T_z=0,\pm1} V^{10}_{ij} 
 P^\dagger_{T_z}(i) P_{T_z}(j) 
\end{eqnarray}
and the results will be labelled as  $T=1$;
  \item p-n pairing: this calculation will illustrate 
qualitatively  the competition between isovector 
  and isoscalar p-n pairing. The isovector p-p and n-n pairing are not considered
  and the following Hamiltonian is employed:
  \begin{eqnarray} 
H& =& H_{s.p.}+\sum_{i\neq j} V^{10}_{ij} 
 P^\dagger_{0}(i) P_{0}(j) 
\nonumber \\
&+&  \sum_{i \neq j} V^{01}_{ij} 
  D^\dagger_{0}(i) D_{0}(j).
  \label{HT0}
\end{eqnarray}
The corresponding SM calculations will be denoted as  $T_z=0$.
\end{itemize}
Denoting by $V$, the two-body part used in the Hamiltonian, a summary of the different cases described above as well as their labels is given in 
 table \ref{Tab:label}.

\begin{table}[h]
\begin{tabular}{ |c|c| }
 \hline                       
  $V$ & label  \\
  \hline \hline
  $ \displaystyle \sum_{i\neq j,T_z=\pm1} V^{10}_{ij} P^\dagger_{T_z}(i) P_{T_z}(j).$ & $\displaystyle |T_z=1|$   \\
  $ \displaystyle \sum_{i\neq j,T_z=0,\pm1} V^{10}_{ij}  P^\dagger_{T_z}(i) P_{T_z}(j)$ & $\displaystyle T=1$  \\
  $\displaystyle \sum_{i\neq j} V^{10}_{ij} 
 P^\dagger_{0}(i) P_{0}(j) 
+  V^{01}_{ij} 
  D^\dagger_{0}(i) D_{0}(j),$&$ \displaystyle T_z=0$\\
  \hline  
\end{tabular}
\caption{Hamiltonian and corresponding labels used in the  paper.}
\label{Tab:label}
\end{table}

\section{Systematic analysis of $T=1$ pairing in N=Z nuclei in the f shell. The constant pairing case.}
\label{Sec:Gconst}
As we will see, the correlations obtained through a direct diagonalization of the pairing Hamiltonian are the results of a subtle 
mixing of different effects like spin-orbit, deformation, beyond mean-field effects, fluctuations in two-body interaction matrix elements. To disentangle the influence of different contributions, we consider below situations of increasing complexity. 
In this section, we only consider $T=1$ pairing. We will in addition first discuss the case of spherical symmetric nuclei with constant two-body matrix elements. The spherical symmetry is imposed 
 for nuclei presented in Fig. \ref{Fig:EvsBeta} by imposing $\beta=0$. The value of the constant pairing strength has been chosen 
 such as to provide a BCS gap approximately equal to 1 MeV calculated at $\beta = \beta_0$, in average for all the considered nuclei. This methodology leads to
 a pairing interaction  $V^{10}= -0.5$ MeV. 
 The diagonalization is made in the restricted valence space formed by the f$\frac{7}{2}$ and f$\frac{5}{2}$ single-particle states (see Figs \ref{Fig:Esp-Fe52} and \ref{Fig:Esp-Ni56} for illustrative examples), and considering the valence nucleons beyond the N=Z=20 saturated core. 

\subsection{Pairing correlations beyond the independent quasi-particle picture}

In mean-field approaches, only n-n and p-p pairing 
 are usually treated self-consistently. Moreover, BCS results are affected by 
 particle number fluctuations, whose quantitative effect can be seen
 by comparing them with the SM calculations, where the particle number symmetry is preserved.   To illustrate the extra correlations included in SM description, 
 we first consider a  $|T_z|=1$ calculation, (see Table \ref{Tab:label}), where only  like particles 
 can interact. 
\begin{figure}[htbp]
\begin{center}
\includegraphics[width = 7.cm]{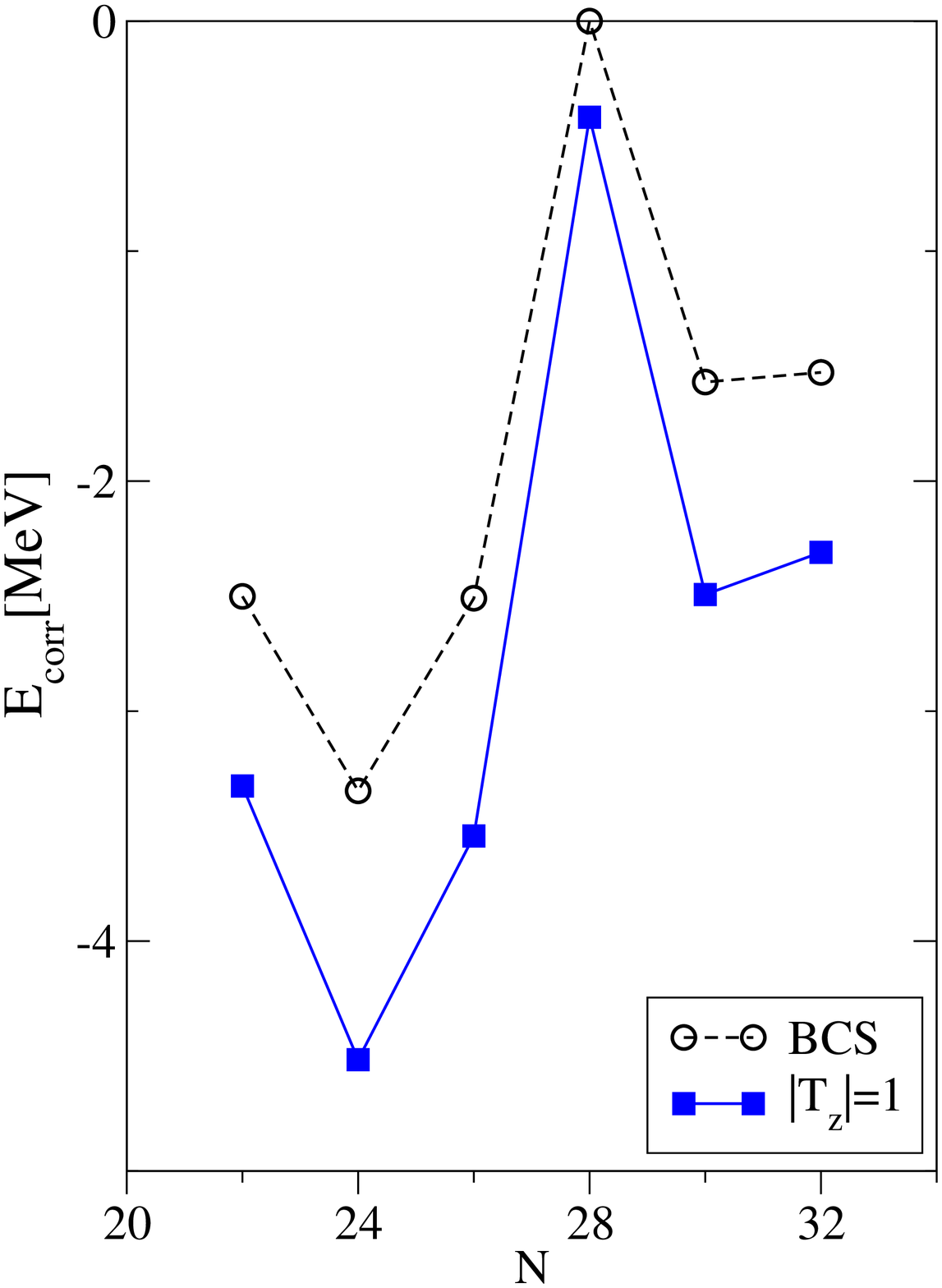}
\end{center}
\caption{(Color online) Correlation energy obtained assuming only pairing between like particles. The BCS (black circles) and $|T_z|$=1 SM results (blue squares) are shown for different $N=Z$ $f$-shell nuclei.
Spherical symmetry is imposed  for all the nuclei and a constant pairing interaction is used (see text).}
\label{Fig:fig1}
\end{figure} 
 In Fig. \ref{Fig:fig1}, the correlation energies obtained in the HF+BCS approximation  with a constant 
$V^{10}$  are shown for different f-shell $N=Z$ nuclei. The correlation energy is defined with respect to the unperturbed case, i.e. $V^{10}$=0.
 These results are compared with the equivalent SM calculation with only $|T_z|$=1 
 pairing correlation. The use of an exact diagonalization compared to the BCS approach has several advantages that are illustrated in this figure. 
 First, the BCS approximation suffers from the so-called BCS-threshold anomaly leading to a zero pairing correlation energy if the single-particle shell-gap is large 
 compared to the residual interaction strength. This is illustrated in the $N=28$ nucleus where the pairing vanishes.  A SM calculation leads to non-zero pairing correlation in this case. In addition, it is known that BCS underestimates the pairing correlation. We see indeed that the pairing correlations obtained with the SM case are sensibly larger in all nuclei.  
 
 \subsection{Competition between particle-like and p-n pairing in spherical nuclei}

Another drawback of mean-field approaches is that, most often it cannot describe the coexistence of pairing in different isospin channels. 
Usually, when p-n symmetry is assumed, BCS or HFB approximations lead to degenerate solutions (either  to a particle-like type condensate or a p-n pair condensate \cite{Gin68, San09,Ber10}).  Such a non-coexistence seems however to be an artifact of the theory. Indeed, it is for instance 
in contradiction with schematic models where exact solution of the pairing problem can be obtained. The full diagonalization of the pairing Hamiltonian, by going beyond the independent picture also cures this problem. 

\begin{figure}[htbp]
\begin{center}
\includegraphics[width = 7.cm]{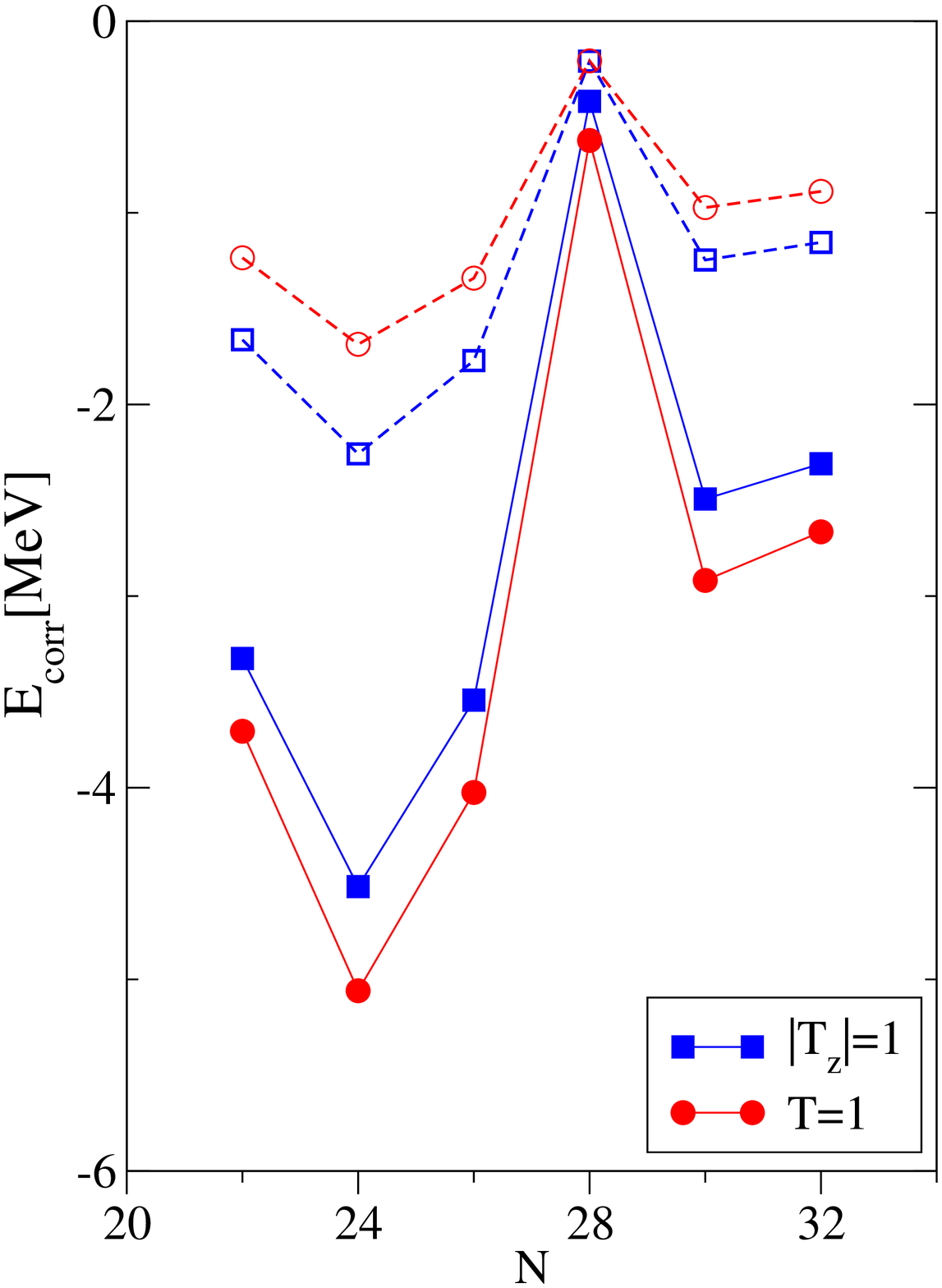}
\end{center}
\caption{(Color online) Correlation energy obtained in SM calculation including all channels in $T=1$ (red filled circles)
compared to the case where only $|T_z|$=1 are included (filled blue squares). In the former case, the red open circles 
indicates the correlation energy associated to one $T_z$ component. Note that due to the assumed symmetry between 
protons and neutrons the three components leads to the same contribution equal to $1/3$ of the total energy. 
The blue open squares correspond to the correlation energy associated to n-n pairs for the $|T_z|$=1 calculation. Again, due to 
the isospin symmetry, this correlation energy is half of the total energy and  equals the one associated to the p-p contribution.  }
\label{Fig:fig2}
\end{figure}

Here, the SM calculation has been performed again in spherical nuclei with constant pairing interaction 
including all channels in the $S=0$ and $T=1$. In particular, the p-n pairing is  accounted 
for in the $T=1$ channel.  Results obtained in this case are compared in Fig.  \ref{Fig:fig2}  with the previous 
SM calculation where only the $|T_z|$=1 pairing was considered. 
 It is seen that the inclusion of the p-n pairing in the $T=1$ slightly increases the total correlation energy.  Since we consider here 
explicitly p-n symmetry, the correlation energy exactly splits into three equals components associated to each $T_z$ 
projection. The correlation energy in one of the $T_z$, that is equal to $1/3$ of the correlation is also show 
by dashed blue lines in Fig.  \ref{Fig:fig2}. This energy is compared to the equivalent correlation energy obtained 
when only particle-like pairing is included ($1/2$ of the total energy obtained in the $|T_z|$=1 case). While the total correlation energy is globally 
increased, we observe that the correlation associated to n-n and p-p pairing is decreased when p-n channel is included compared to 
the case where it was neglected.  The interpretation of this quenching is rather intuitive. When a neutron (resp. a proton)
does already contribute to a pair with a proton (resp. a neutron), it cannot anymore be used to form a pair with a neutron (resp. a proton). This effect 
can therefore be understood as an indirect pair breaking effect induced by the addition of an extra channel.

\begin{figure}[htb]
\begin{center}
\includegraphics[width = 7.cm]{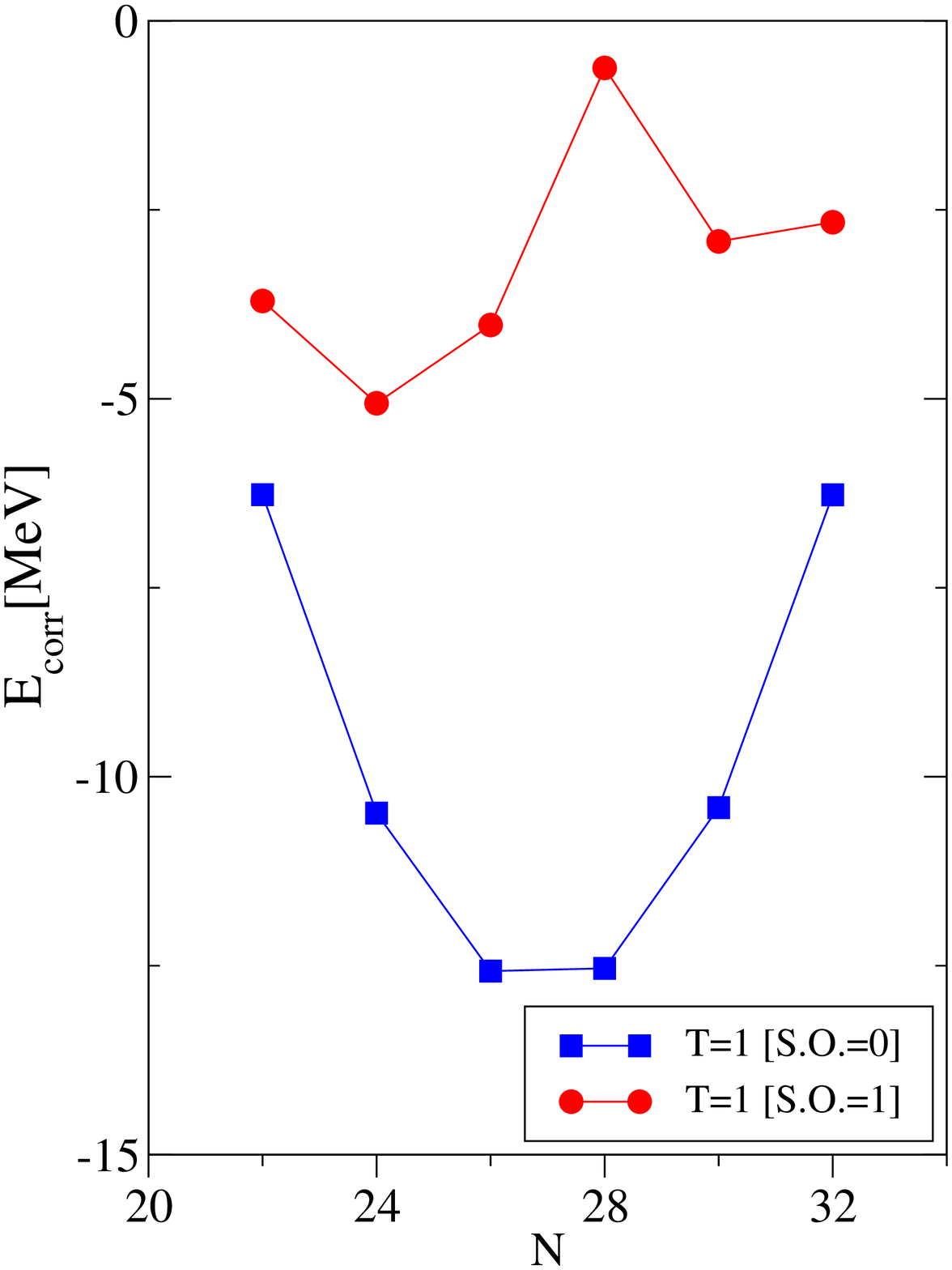}
\end{center}
\caption{(Color online) Correlation energy in the T=1 case
with (red filled circles) and without (blue filled squares) spin-orbit interaction in the mean-field calculations.
The results are obtained assuming a spherical symmetry for all nuclei and a constant pairing interaction.}
\label{Fig:fig3}
\end{figure} 

\subsection{Spin-orbit effect in spherical nuclei}
Due to its effect on single-particle energies, it is known that the spin-orbit interaction affects globally pairing. As a quantitative illustration of the spin-orbit influence, 
we compare in Fig. \ref{Fig:fig3} two $T=1$ SM calculations. The first one (red full circles) is the one described above and includes the effect of spin-orbit 
on single-particle energies with the presence of the $N=28$ magic number. In the second one (blue filled square), the spin-orbit interaction
has been artificially set to zero, leading to a completely degenerated $f$ shell.  This figure gives quantitative information on the reduction resulting from the spin-orbit interaction and will serve in the following as an element of comparison for the effect of deformation. 
 
\subsection{Deformation effect: constant pairing case} 
 
\begin{figure}[htb]
\begin{center}
\includegraphics[width = 8.cm]{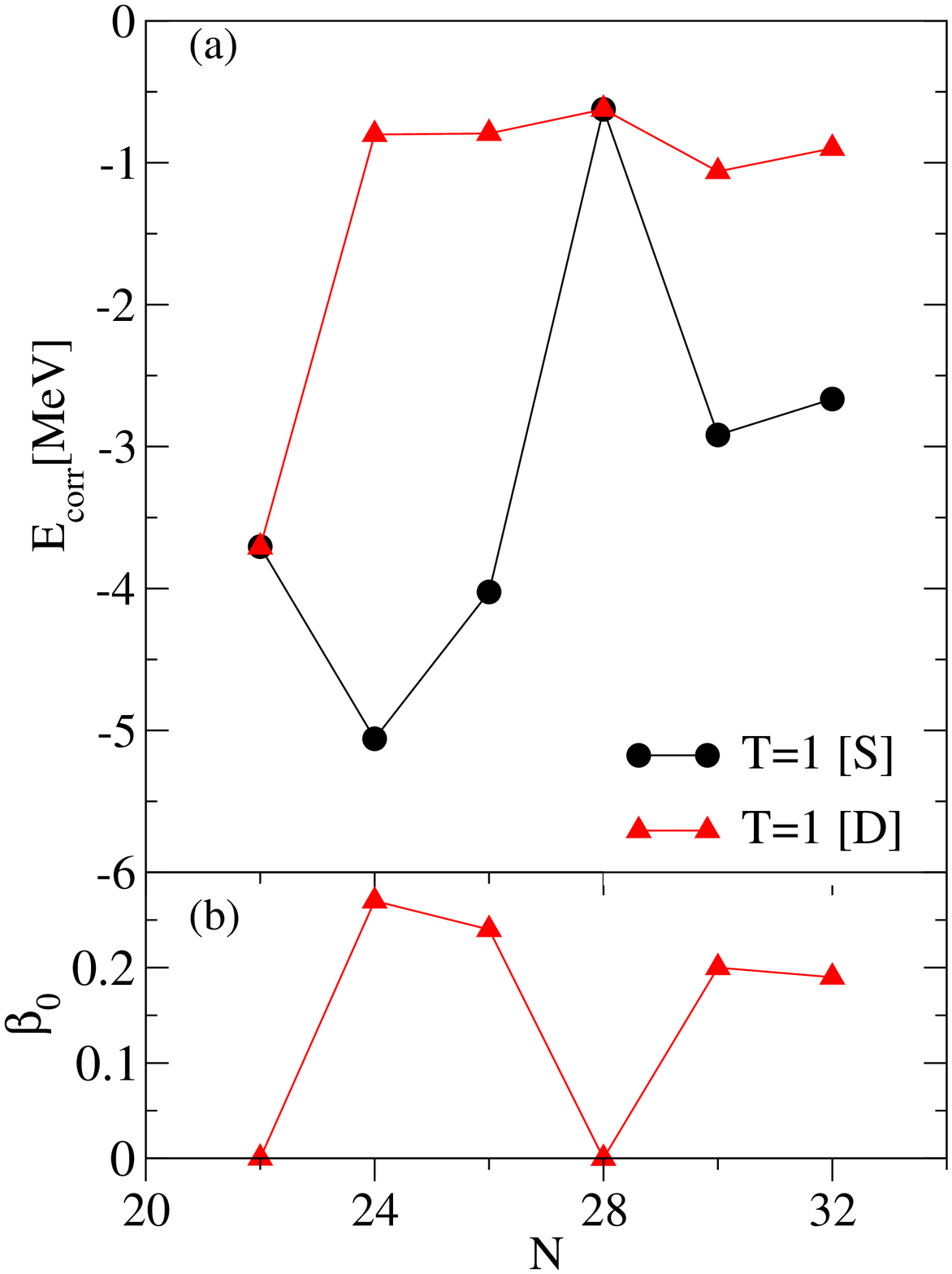}
\end{center}
\caption{(Color online) a) Correlation energy in the $(S=0,T=1)$  obtained with SM calculation in spherical (black filled circles)
and deformed (red filled triangles) $N=Z$ nuclei. In the latter case, the equilibrium deformation value $\beta_0$ obtained with {\sl EV8} 
was used. b) value of  $\beta_0$ deduced from Fig. \ref{Fig:EvsBeta}  as  a function of the neutron number. }
\label{Fig:fig4}
\end{figure}

To conclude the set of calculations with constant pairing interaction, we also performed SM calculation for deformed cases. 
For all considered nuclei, in Fig. \ref{Fig:fig4}-a, we have systematically compared the correlation energy in spherical configurations (black symbols) with the results obtained when the 
deformation is equal to the equilibrium deformation value $\beta_0$ (red symbols). The $\beta_0$ values correspond to the minima of the energy landscapes displayed in Fig. \ref{Fig:EvsBeta} and are reported in Fig.  \ref{Fig:fig4}-b. 
Compared to the spin-orbit, we can see that deformation has a less impressive effect. However, in all 
considered nuclei, a significant reduction of the $T=1$  correlation energy is observed 
in deformed nuclei compared to the reference spherical case. Since the interaction is kept constant 
here, this reduction is a direct consequence of the single-particle shell evolution. Such a reduction 
was observed in most cases already at the mean-field level. For instance,  in Fig. \ref{Fig:GapvsBeta}, 
in all non-closed shell nuclei except $^{64}$Ge, the pairing gap in the spherical configuration is always larger than the one in the equilibrium 
deformed configuration.
When SM calculations are performed instead of HF+BCS, the effect of deformation seems 
much stronger. In all deformed nuclei, the correlation energy is reduced by more than 60 \%.  This is an important effect that 
points out that deformation plays a major role that should definitively be included in quantitative studies.  It is finally worth to mention that the effect of $N=28$, that was clearly seen in spherical nuclei, is almost completely washed out 
when deformation is accounted for. The calculation with deformation leads to an almost constant, rather weak, $T=1$ pairing correlation 
from $N=24$ to $N=32$. 

As discussed above, the results shown in this section are obtained using quite strong approximations on both the interaction and valence space and only qualitative conclusion can be drawn from them. We analyzed here the interplay between different effects, i.e. pairing, deformation and spin-orbit. 
In the following section, we will see that these effects persists even if a more realistic analysis is made using 
both improved two-body pairing interaction and a valence space authorizing the mixing of $f$ and $p$ single-particle levels.

\section{Results with realistic residual interactions}
\label{Sec:Greal}
In the previous section, deformation effects were considered only in the
single-particle energies employed as input quantities in the subsequent SM calculations. Moreover,
only the T=1 channel has been considered in the residual interaction. A more realistic and consistent description can be reached by calculating the two-body pairing matrix elements for
the interaction (\ref{eqn:vpair}) consistently with the mean-field properties.  In such a way,  two-body pairing matrix elements are affected by deformation through its effect on the single-particle states topology. In addition, by using standard projection techniques as explained in Appendix \ref{app:2body}, the two different channels of the pairing interactions can be accounted for and 
the interplay between p-n isovector and isoscalar correlations analyzed. This is a merit of the strategy used in this work, since common studies employing constant pairing strength interactions
in the two channels are not able to describe important effects as for example the stronger
quenching  of the isoscalar matrix elements with respect to the isovector ones due to the spin-orbit interaction\cite{Sag13,Poves1998,Baroni2010}.

\subsection{Effect of deformation on the residual interaction}
\begin{figure}[htbp]
\begin{center}
\includegraphics[width = 7.cm]{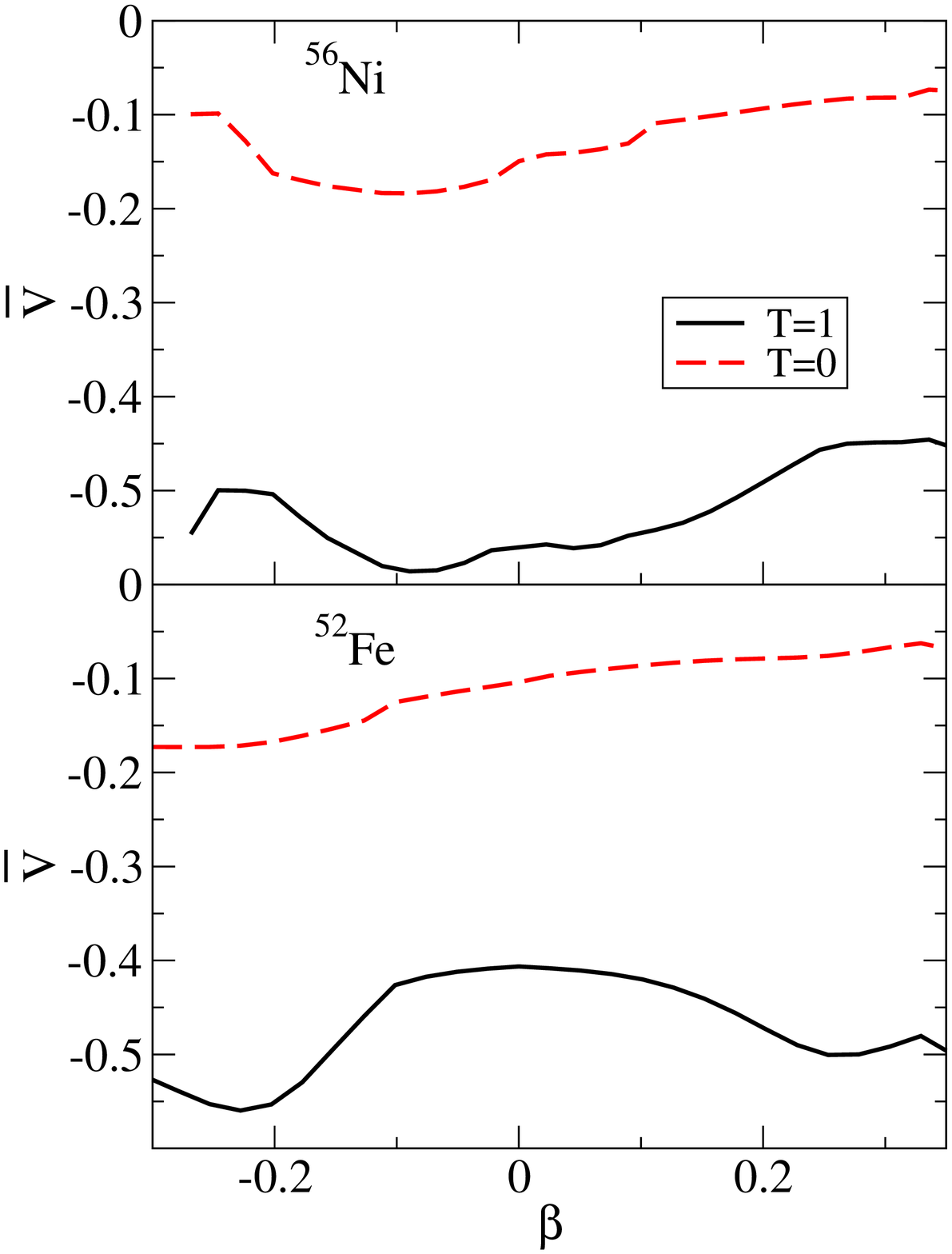}
\end{center}
\caption{(Color online) Average of the pairing matrix elements in the T=1 (black solid line) and T=0 (red dashed line) channels
as a function of the deformation for $^{56}$Ni (upper panel) and  $^{52}$Fe (lower panel) are plotted.
The diagonal part of the interaction is not considered in the calculations and not shown here (see the text for more details). The matrix elements are 
computed here using the same strength in the two channels, i.e. $v_1=v_0$ ($x=1$) parameter in Eq. (\ref{eqn:vpair}). }
\label{Fig:Gmat1}
\end{figure}

In order to analyze the role that deformation can have on the residual interaction, the evolution of the average $T=1$ two-body matrix elements (black solid line) and $T=0$ ones (red dashed line)  obtained between 4 particles and 4 holes around the Fermi energy are shown in  Fig. \ref{Fig:Gmat1} as a function of $\beta$ for the two nuclei   $^{56}$Ni (upper panel) and  $^{52}$Fe (lower panel). 

\begin{figure}[htbp]
\begin{center}
\includegraphics[width = 8.cm]{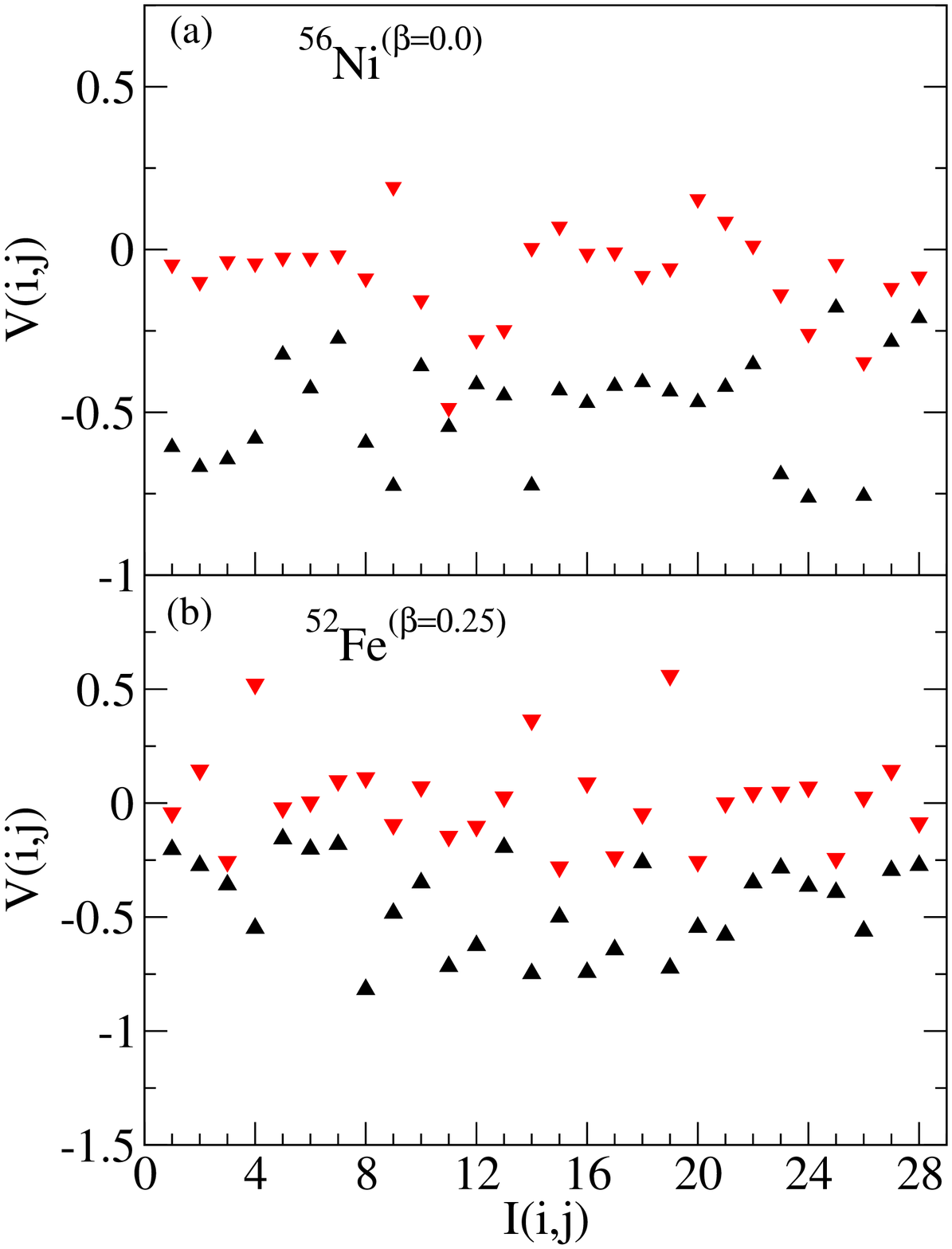}
\end{center}
\caption{(Color online) Pairing matrix elements in the T=1 (black points) and T=0 (red points) channel
for 8 single-particle states (4 hole and 4 particle states)  around the Fermi energy
as a function of the deformation for $^{56}$Ni (upper panel) and  $^{52}$Fe (lower panel) are plotted.
The index $I(i,j)=1,..,28$  enumerates the pair indices of the matrix elements $V_{ij}$
with $i=1,..,8$ and $j>i$. 
The wave functions used corresponds to the $\beta$ equilibrium 
value. The matrix elements are 
computed here for $v_1=v_0$ ($x=1$). }
\label{Fig:Gmat2}
\end{figure}

Note that, we do not include in the average and in the SM calculations the diagonal part of the matrix elements $V^{10}_{ii}$. 
Indeed, including this part would induce a double-counting of the interaction due to the mean-field term entering implicitly 
in the calculation of the single-particle energies.  
It is finally worth to mention that the calculated two-body matrix elements are far from the constant coupling limit especially in deformed nuclei. 
In Fig. \ref{Fig:Gmat2}, the fluctuations of pairing matrix elements in the T=1 (black points) are shown
for $^{56}$Ni (upper panel) and  $^{52}$Fe (lower panel). In each case, the configuration shown is the equilibrium value, i.e. spherical for the 
 $^{56}$Ni and deformed with $\beta= 0.25$ for  $^{52}$Fe. 

\subsection{Comparison between constant and realistic pairing interaction}

In all SM calculations presented previously, the interaction was taken constant 
with a strength compatible with the average strength obtained using realistic residual 
interaction (see Fig. \ref{Fig:Gmat1}). In spherical nuclei, top panel of Fig. \ref{Fig:Gmat2}, the fluctuations 
of two-body matrix elements around the average values are rather small. In deformed nuclei, the fluctuations 
seem to be enhanced. This behavior might directly be attributed to the effect of deformation on the spatial 
localization of the single-particle wave-packet. This can further modify the pairing correlation itself.  In the following, we use the state dependent 
matrix elements of the pairing
interaction in the two different channels (see appendix \ref{app:2body}).

As already discussed, a rather strong dependence of the single-particle states
as a function of the deformation comes out from the mean-field description, (see Fig. \ref{Fig:Esp-Fe52} and \ref{Fig:Esp-Ni56}). Moreover, since in the rotational symmetry is explicitly
broken their identification through  angular quantum numbers is not possible. To study the
effect of deformation through all the energy landscape we consider now the following strategy. SM calculations are performed for each nucleus and for each deformation  considering 8  (proton and neutron) single-particle states (4 hole and 4 particle states)  around the Fermi energy and 8 valence particles for each kind of nucleons. This amounts to adopt a pairing window around the Fermi energy of about 5 MeV, which is consistent with the one employed in the {\sl EV8} calculations.

\begin{figure}[htb]
\begin{center}
\includegraphics[width = 8.cm]{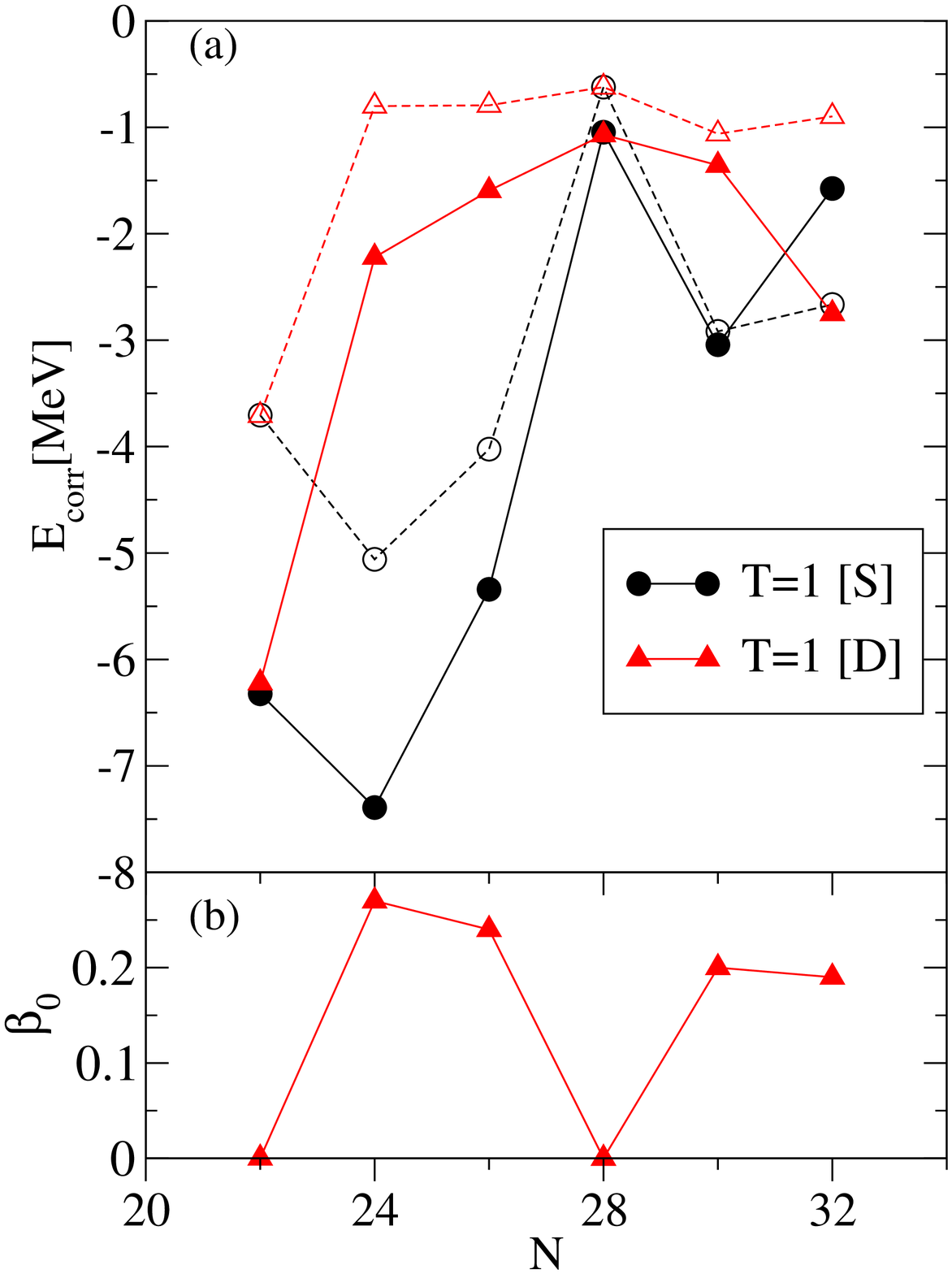}
\end{center}
\caption{(Color online) 
Correlation energies obtained in T=1 SM calculations with residual pairing interaction computed directly from {\sl EV8}. For deformed nuclei, see panel (b), the results obtained using the spherical
mean-field {\sl EV8} solution (filled black circles) and the deformed one (filled red triangles) are shown. The results obtained by using a constant pairing interaction (open symbols) are also shown for comparison.}
\label{fig:4bis}
\end{figure}

In Fig. \ref{fig:4bis}, the correlation energies reported for a constant coupling in Fig.  \ref{Fig:fig4} are systematically 
compared with the SM result where the two-body matrix elements are calculated
from the {\sl EV8} outputs.   
\begin{figure*}[htb]
\begin{center}
\includegraphics[width = 16.cm]{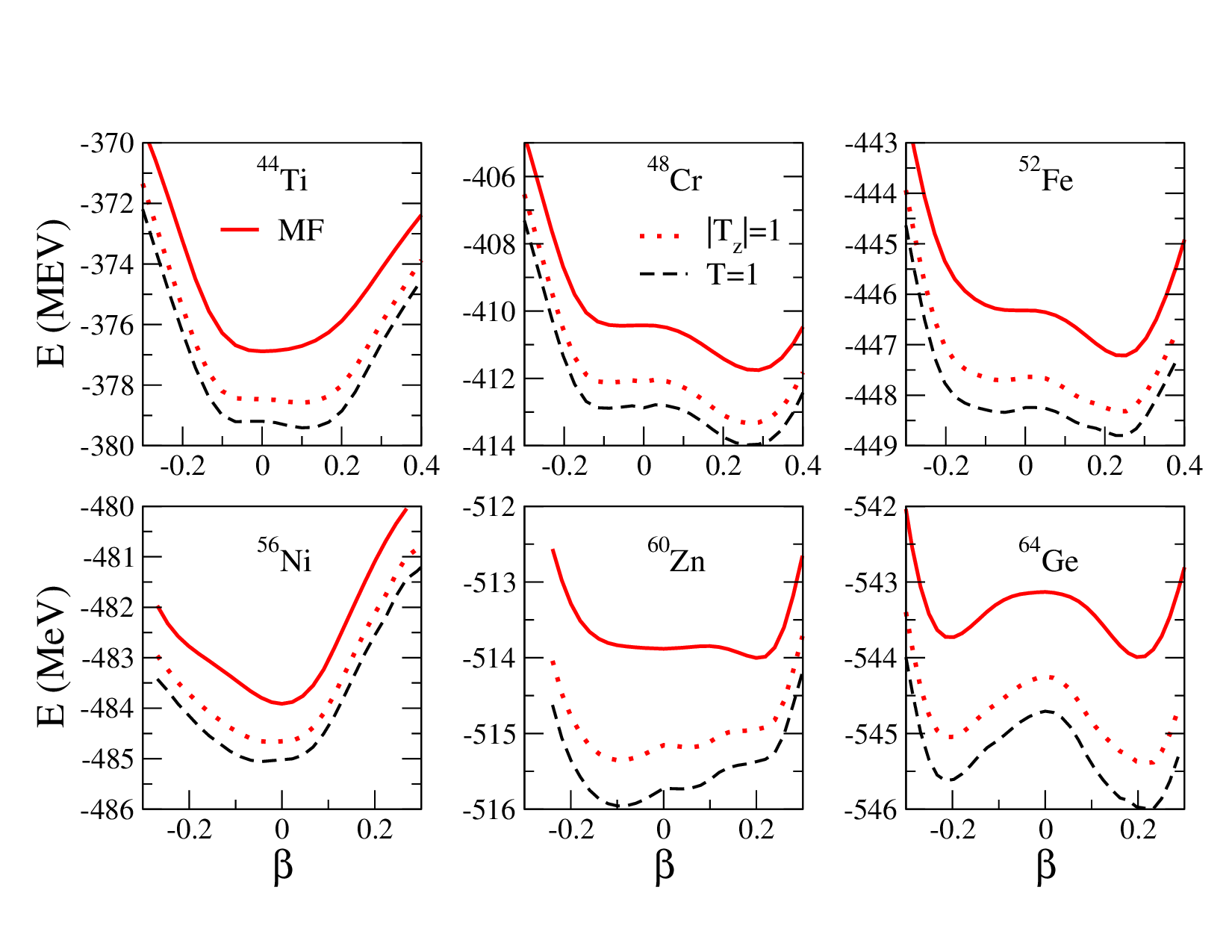}
\end{center}
\caption{(Color online) The total energy (\ref{Eq:ETOT}) as a function of the deformation parameter calculated  in the mean-field (MF) approach is compared with different kind of SM calculations including T=1 correlations. See the text and Table \ref{Tab:label} for more details. 
 }
\label{Fig:EtotvsBeta}
\end{figure*}
Several interesting aspects can be seen in this figure. First, the flattening of the correlation  energy associated to the reduction of pairing induced by deformation around  the $^{56}$Ni is also seen. This effect seems to be generic and does persist even if the residual interaction also account for 
deformation. Besides this global effect, we also observe mainly two differences between the new results and the constant interaction case. 
First, the correlation energy is slightly increased compared to the previous case. This might appear surprising due to the fact that the constant 
interaction was set to a value $0.5$ MeV that is consistent with the average interaction of the $T=1$ channel reported in Fig. 
\ref{Fig:Gmat1}. However, Fig. \ref{Fig:Gmat2} points out that important fluctuations of the two-body matrix 
elements exists especially in deformed systems. The second important difference is the $^{64}$Ge where the new calculation leads to a
significant increase of the pairing correlations compared to the previous case. This directly stems from the valence space ($f$ shell) used 
previously that was too restrictive to built up correlation in this nuclei. In the new results, the particles can scatter to the $p$-shell leading 
to a more realistic description of pairing.        

\subsection{Global study of deformation and $T=1$ pairing on the energy landscape}

Although the correlation energy can easily be obtained from SM calculation, a complete potential energy landscape 
deduced by combining a mean-field and a SM calculation is less straightforward. In particular, since {\sl EV8}
is a fully self-consistent calculation it already contains the effect of pairing in both the mean-field and anomalous contribution.

To construct an energy landscape associated to the SM results, we follow the strategy of
Ref. \citep{Alh2011}. We systematically performed an equivalent BCS calculation using 
the same {\it EV8} inputs and compute the energy by subtracting the BCS pairing part to the energy using the formula  
 \begin{equation}
  E=E_{\sl MF}+E_{SM}-E_{BCS}.
  \label{Eq:ETOT}
 \end{equation} 
 Here $E_{\sl MF}$ is the total (mean-field + pairing) energy directly obtained from {\it EV8}. $E_{SM}$ is the total SM energy 
 deduced from the calculation and $E_{BCS}$ is the BCS energy calculated by solving the BCS equations 
using the same inputs and constraints of the SM case (i.e sp states, active particles, and Hamiltonian (\ref{Eq:plikeH})).
\begin{figure*}[htb]\begin{center}
\includegraphics[width = 16.cm]{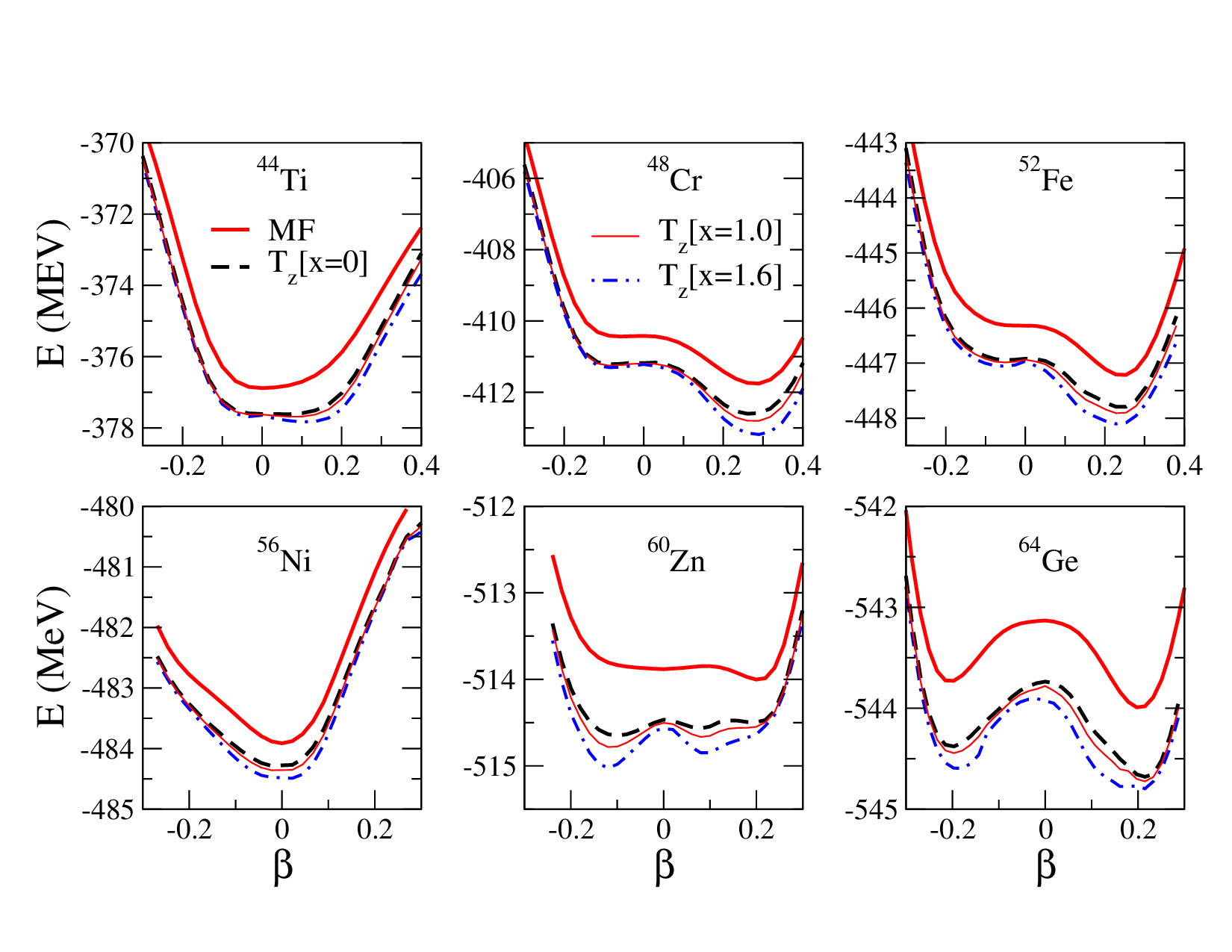}
\end{center}
\caption{(Color online) Total energy (\ref{Eq:ETOT}) as a function of the deformation obtained using only p-n pairing in the T=1
channel (black dashed line, $T_z[x=0]$), including also the isoscalar one with equal strength ,(thin-solid red line, $T_z[x=1]$) and  using the  more realistic value $x=1.6$ (dot-dashed blue line, $T_z[x=1.6]$). The  results obtained in the mean-field (MF) calculations are plotted for comparison.} 
\label{Fig:EtotvsBeta2}
\end{figure*}
In Fig. \ref{Fig:EtotvsBeta}, the energy $E$ computed from Eq. (\ref{Eq:ETOT}) is shown for the 6 considered nuclei as a function of the deformation 
parameter. The results obtained in the $(S=0,T=1)$ channel are shown either including all channels or only particle like pairing.  
In all cases, the results are compared with the energy landscape obtained directly from {\sl EV8}. 

Considering first the SM model results with only particle like pairing, i.e.  $|T_z|=1$, we see  that additional correlation energy is systematically gained compared to the original {\sl EV8} case, due to the use of a complete diagonalization in a particle conserving SM approach. In most cases, the better treatment 
of pairing leads to a global shift of the energy landscape by few ($\simeq 2$) MeVs. In the case of $^{60}$Zn, a transition from prolate to oblate shape 
occurs in the energy landscape.  
It should be noted that the original BCS energy landscape in that case was rather flat over a wide range of deformation  parameter values although a prolate minima was predicted. 
It turns out that the preservation of the particle number symmetry, which is spontaneously broken in BCS, allows 
to gain some extra correlation energy. For $^{60}$Zn, this energy gain is bigger in the oblate configuration compared to the prolate case leading to  
the observed transition. In the same figure, the SM results obtained considering also the p-n pairing in the $T=1$ channels are plotted in dashed-black line. Compared with the $|T_z|=1$ results, we see that the inclusion of p-n correlations provide some extra correlation energy, typically of the order of 
few hundreds of KeV. Also in this case, the effect is more pronounced for $^{60}$Zn, and the p-n correlation seems to enhance the transition from prolate to oblate.  


\begin{figure*}[htb]
\begin{center}
\includegraphics[width = 16.cm]{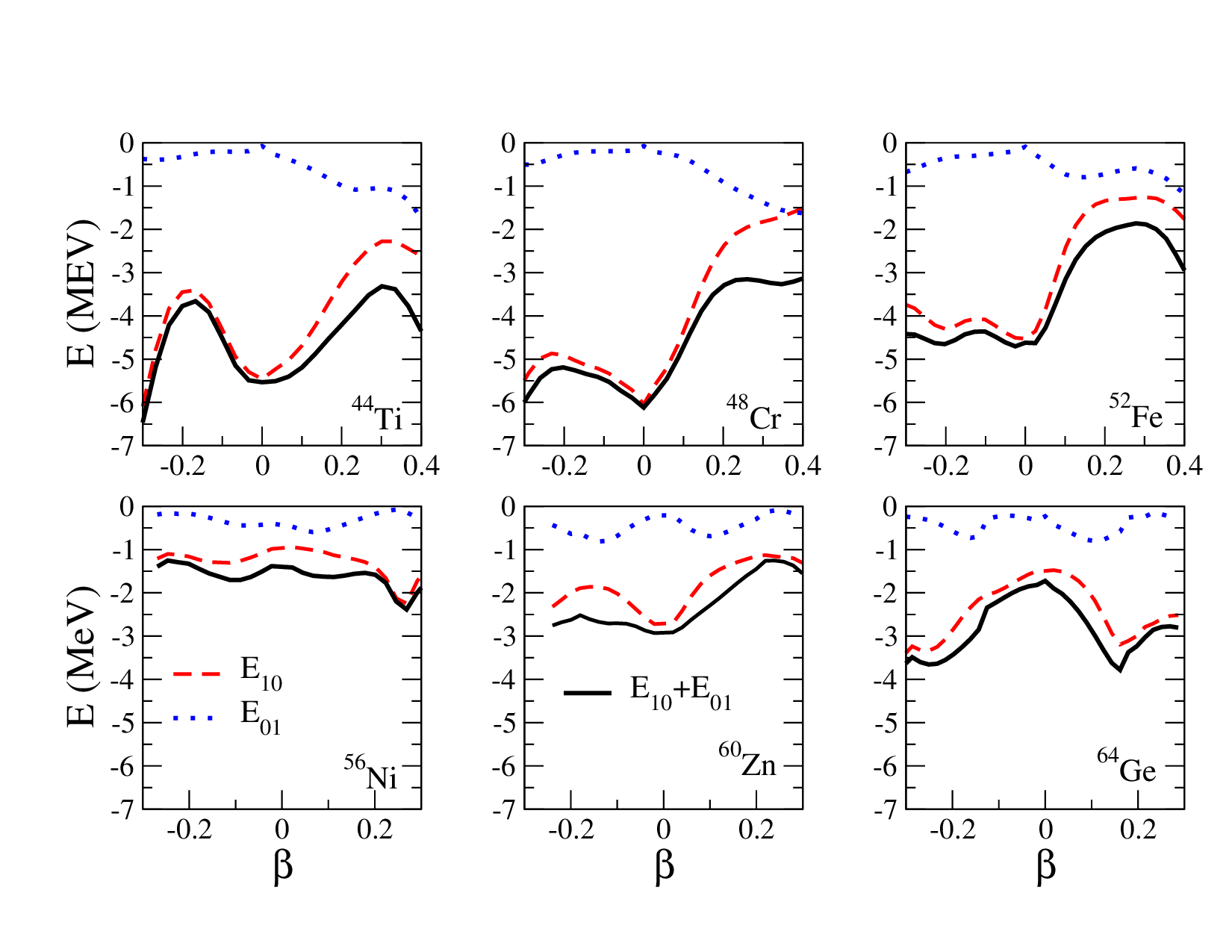}
\end{center}
\caption{(Color online)  The isovector (red dashed line)  and isoscalar (blue dotted line)  energy contributions, defined by Eqs (\ref{Eq:MeanV10}) and
(\ref{Eq:MeanV01}) respectively, and the sum of them (black solid line) are plotted as a function of the deformation and corresponding to a ($|T_z|=0$) calculation with $x=1.6$.}  
\label{Fig:Etopercent}
\end{figure*}

\subsection{Competition between $T=0$ and $T=1$ proton-neutron pairing}

As we mentioned previously, only the $S_z=0$ component of the $T=0$ channel can use the direct output from 
the ${\sl EV8}$ code. Here, we will concentrate on the p-n pairing correlation and study its competition in the two pairing channels. Therefore, the pairing between particles of same isospins is neglected.  To calculate the interaction 
matrix elements in the isoscalar case, we follow Refs. \cite{Ber10, Leb12} and use the same residual interaction
(\ref{eqn:vpair}) where $v_0$ is replaced by $v_1=x v_0$. The resulting average interactions as well as fluctuations around the average 
are shown in Fig. \ref{Fig:Gmat1} and \ref{Fig:Gmat2} respectively for $x=1$. We already see in these figures, that the 
interaction matrix elements in the $T=0$ channel are systematically smaller compared to the $T=1$ case. This quenching has been 
already studied and can be traced back to a stronger
effect of the spin-orbit interaction in the T=0 case, see for example \citep{Sag13,Poves1998,Baroni2010}. It has  been mentioned above that a realistic value of $x$ is 1.6. This value will only partially compensates the fact that higher pairing interaction strength
exists in the isovector channel. 
In Fig.  \ref{Fig:EtotvsBeta2}, the potential energy curve obtained including only isovector 
p-n pairing (dashed-black line), i.e. x=0, is compared with the results where isoscalar
correlation are included also, for different x values.  It is clear from this figure that the addition of the isoscalar channel gives a gain in correlation energy that is almost negligible. It is however worth mentioning 
that this does not necessarily means that the two channels do not mix. Using the same argument as before, i.e. that isovector p-n pairs can be broken 
to form isoscalar p-n pairs, the reduction of isovector pairing energy would eventually be compensated by an increase of the isoscalar 
pairing energy. We also see, comparing the results with $x=1.0$ and $x=1.6$ that the dependence
of the isoscalar correlation energy on the $x$ value is  rather weak.
\begin{figure*}[htb]
\begin{center}
\includegraphics[width = 16.cm]{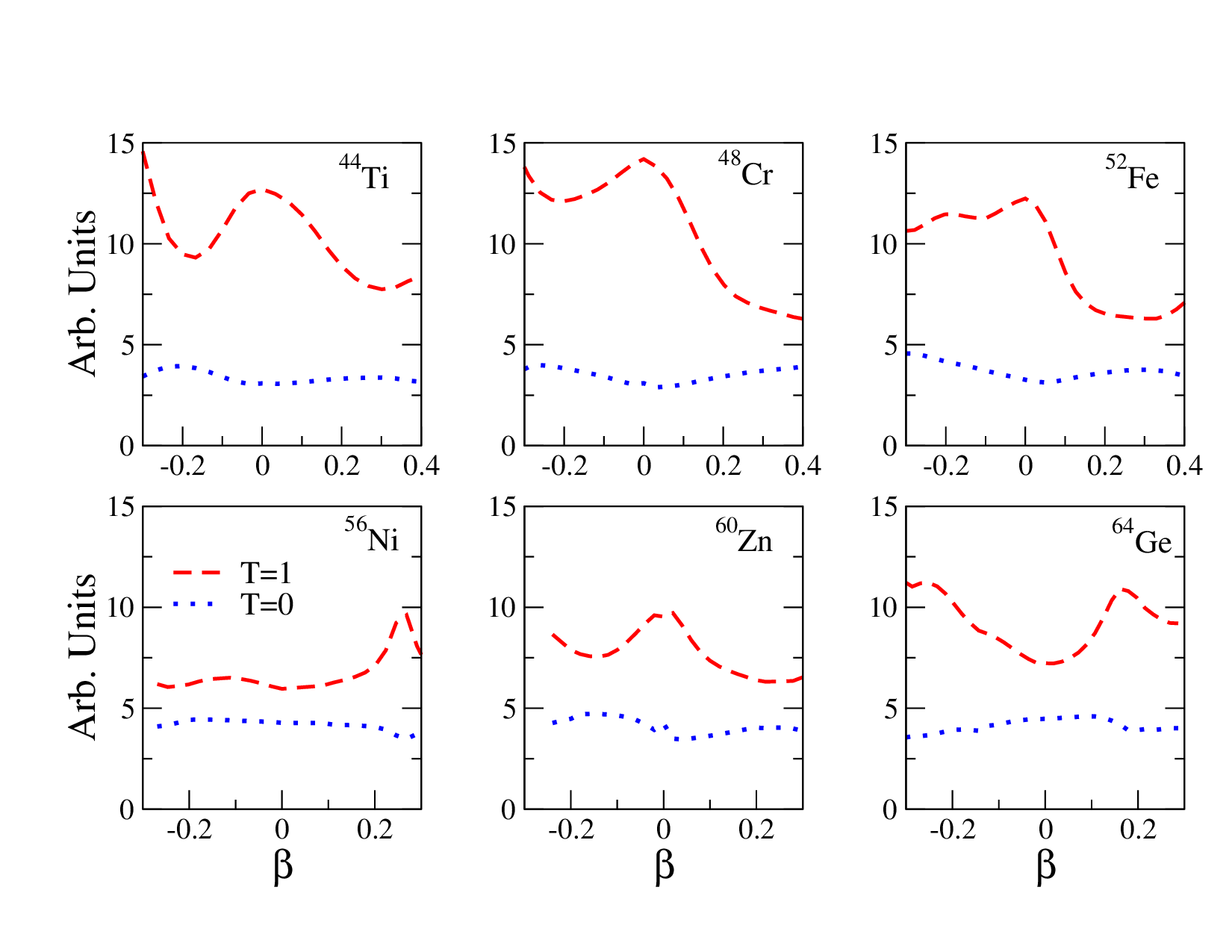}
\end{center}
\caption{(Color online) Isovector (red solid line) and isoscalar (blue dashed line) total deuteron transfer probability obtained for $x=1.6$, corresponding,respectively, to the $Q$ and $R$ quantities defined in the text.}
\label{Fig:ptrans}
\end{figure*} 
To study the competition between isoscalar and isovector correlations, we plot in Fig.
 \ref{Fig:Etopercent} the expectation values of the two-body interactions corresponding to the two different channels appearing in the Hamiltonian (\ref{HT0}) evaluated in the ground state obtained in a $T_z=0$ calculation. More precisely, we calculate the following quantities:
 \begin{equation}
E_{10} =\langle \sum_{i\neq j} V^{10}_{ij} 
 P^\dagger_{0}(i) P_{0}(j)\rangle, 
 \label{Eq:MeanV10}
 \end{equation}

\begin{equation}
E_{01} =\langle \sum_{i\neq j}V^{01}_{ij} 
  D^\dagger_{0}(i) D_{0}(j) \rangle. 
  \label{Eq:MeanV01}
 \end{equation} 
 From the figure we cannot see a common and regular dependence on the deformation of these quantities. However, for all nuclei, the expectation value of the isoscalar Hamiltonian
 is generally smaller than the isovector one, although a value $x=1.6$ has been used. 
For the $^{48}$Cr nucleus and to a less extend for the $^{52}$Fe nucleus, 
we see that at equilibrium deformation values, the two pairing channels 
 (isoscalar and isovector) contributes significantly showing the possible coexistence 
 of both pairing. While absent in most quasi-particle approaches to p-n pairing except in some 
 extreme cases in the nuclear chart \cite{Gez11}, a SM approach always leads to a non zero pairing correlations 
 in both channels. However, we see in Fig.  \ref{Fig:Etopercent} that the isoscalar energy is rather weak and in most cases, almost 
 cancels out at $\beta=0$. We have however here a clear evidence that deformation can favor in some cases the coexistence 
 of isoscalar and isovector p-n condensates.

 \subsection{Competition of isoscalar and isovector pairing on deuteron transfer}

In this section, we still focus on the $T_z = 0$ calculation.
The role of different pairing channels on the deuteron transfer cannot directly be 
inferred from the energy consideration given above. To get information on this aspect, we
consider the two, respectively, isovector and isoscalar operators:  
\begin{eqnarray}
  \hat{Q}_{10} &=&  \sum_{ij} \hat{P}^\dagger_{0}(i)  \hat{P}_{0} (j),  \nonumber \\
  \hat{R}_{01} &=&  \sum_{ij} \hat{D}^\dagger_{0}(i)  \hat{D}_{0} (j). \nonumber
\end{eqnarray} 
These two operators differ from the two-body part of the $T_z=0$ hamiltonian by the absence of the two-body interaction.

 Starting from  the ground state wave-function of a nucleus with $N$ neutrons and $Z$ protons, denoted generically 
 $| N, Z \rangle$, the expectation value of the operator  $  \hat{Q}_{10} $ is given by 
\begin{eqnarray}
Q &\equiv& \langle N, Z |\hat{Q}_{10} | N, Z \rangle \nonumber\\
&=&\sum_\alpha |  \langle  N+1 , N+1, \alpha | \sum_i \hat{P}^\dagger_{0}(i) | N,Z \rangle |^2
\nonumber \\
&=& \sum_\alpha |  \langle  N-1 , N-1, \alpha | \sum_i \hat{P}_{0}(i) | N,Z \rangle |^2. \nonumber 
\end{eqnarray}
where we have introduced a complete basis of the nucleus $(N-1,Z-1)$ or $(N+1,Z+1)$, with states labelled by $\alpha$. 
A similar expression can be obtained for $R =  \langle N, Z |\hat{R}_{01} | N, Z \rangle$. Such expression clearly demonstrates 
that the quantity $P$ or $Q$ gives global quantitative information on the probability to transfer or remove a deuteron from 
the initial ground state.  

In Fig.  \ref{Fig:ptrans}, these two quantities are shown as a function of the deformation parameter.  Comparing Fig. \ref{Fig:ptrans}
and Fig. \ref{Fig:Etopercent}, not surprisingly, we observe a strong correlation between the deuteron pair transfer probabilities and
the corresponding correlation energy in a given channel. In all considered nuclei and whatever is the deformation, the isovector channel
always dominates over the isoscalar channel. However, in some specific cases, again due to the quenching of the isovector pairing with 
deformation, the two contributions might start to compete. 

We mention that the quantities  $R$ and $Q$ besides containing a compacted information on the pair transfer probabilities, they also provide the average value of the p-n pairs number in the T=1 and T=0 channels. Indeed, in the absence of correlations, single-particle occupation numbers are equal to $1$ or zero. 
Then, since we consider 8 valence  nucleons for each type, $Q=R=4$ in the absence of correlations. We see that the value of $R$ is very close to this limit even if correlations are plugged in, showing again the quite weak role of the isoscalar pairing.


\section{conclusion}\label{Sec:concl}

In the present work, we investigate particle-like and p-n correlations and their dependence on nuclear deformation. This is done in a framework that combines
self-consistent mean-field calculations and the full diagonalization of the Hamiltonian
containing both the T=0 and T=1 pairing channels. The self-consistent mean-field calculations
provide the main ingredients, single-particle shells and residual two-body matrix elements, that are used in the subsequent SM calculations. In particular, in such a way, deformation effects are realistically and microscopically described through the Skyrme-HF+BCS self-consistent calculations. The  SM like approach corresponds to a framework  beyond  the independent quasi-particle picture. It has the advantage to explicitly conserve the number of neutrons and protons. The resulting approach can give a precise description of pairing correlations and eventually treat the coexistence of different condensate formed of pairs with different spin/isospin. 
This framework is here used to systematically investigate  fp-shell even-even $N=Z$ nuclei, from $^{44}$Ti to $^{64}$Ge.   We found that, in addition to the important spin-orbit effects, deformation plays also an important role. When isovector pairing only is included, we observe that deformation can lead to a quenching of the pairing correlations compared to the spherical case. This quenching is particularly visible around $N=Z=28$ and tends to wash out the pronounced effect of this magic number  that was observed at the BCS level.   This behavior originates mainly from the evolution of single-particle shell energies with deformation (i.e. appearance or disappearance of shell gaps).
On the other hand, the possibility to calculate the matrix elements of the residual interaction 
in the two different channels consistently with the mean-field solution allows to study
in a realistic way (with respect to the case when schematic constant pairing interactions are employed) the interplay of the isoscalar and isovector correlations, and their quantitative role
in the binding energies. It is found that the isoscalar p-n pairing is generally much weaker than the isovector contribution. However, in some cases, we observed that large deformation  can favor the  coexistence  of isoscalar and isovector p-n condensates. We finally analyzed the competition of isoscalar and isovector p-n pairing and its possible influence on deuteron pair transfer.
Also in this case, it is found that the isoscalar  p-n pairing is generally much weaker than the isovector one.  Recently, experiments aiming at disentangling both origins of p-n pairing have been proposed. Although we found that isovector pairing effects is small 
on deuteron transfer, it should however be kept in mind that, experimentally, the transfer associated 
to isovector or isoscalar correlation can \textit{a priori} be measured separately since they correspond to different total spin transfer channels. Therefore, with the ratio of pair transfer probabilities one could  removes the part of the uncertainty on the relative strength of the residual isoscalar and isovector interactions, i.e. the $x$ parameter value. In this respect,
pair transfer probabilities  corresponding to the different channels can be calculated using the many-body wave functions (of ground state and low lying excited states) obtained with the present framework. Work in this direction is in progress.
 
\section{Acknowledgments }
D.G.  thanks P. Van Isacker for many fruitful discussions at the initial stage of this work.
  \begin{widetext}
 \appendix
 \section{Matrix elements of the pairing Hamiltonian}
 \label{app:2body}
 Let's consider the single-particle state  labeled by the index $k$\footnote{Note that for
a deformed nucleus, the single-particle wave functions
are in general not eigenstates of the angular momentum operators.} and by the isospin quantum
 number $\tau_k$ 

 \begin{equation}
 |k \tau_k \rangle= \int d^3 r\sum_{\sigma_k} \phi_k(\sigma_k,r) |r \sigma_k \tau_k\rangle
\end{equation} 
where $\phi_k(\sigma_k,r)$ is the single-particle wave function with spin projection $\sigma_k$.
We can introduce the two body states 
 \begin{equation}
 |k\tau_k,\bar{k}\tau_{\bar{k}}\rangle= 
  \int d^3 r_1 d^3 r_2 \sum_{\sigma_k,\sigma_{\bar{k}}}
 \phi_k(\sigma_k,r_1) 
  \phi_{\bar{k}}(\sigma_{\bar{k}},r_2) |r_1 \sigma_k \tau_k;r_2 \sigma_{\bar{k}} \tau_{\bar{k}}\rangle.
 \end{equation}
 For a given interaction $V$ we can define the corresponding 
two-body matrix elements in the spin-isospin channels as
 \begin{equation}
 \langle i \tau_i  \bar{i} \tau_{\bar{i}}    |V^{T, S}| j \tau_j  \bar{j} \tau_{\bar{j}}\rangle=
 \langle i \tau_i  \bar{i} \tau_{\bar{i}}|VP_{S}P_{T}| j \tau_j  \bar{j} \tau_{\bar{j}}\rangle
 \nonumber
\end{equation} 
where $P_{S},P_{T}$ are the  standard  spin-isospin projection operators
 \begin{eqnarray}
P_S  &= & \frac{1}{2} (1 - (-1)^S P_\sigma) ,~~~
P_T =   \frac{1}{2} (1 - (-1)^T P_\tau) \nonumber
\end{eqnarray}
where $P_\sigma$ and $P_\tau$ exchange the spin and isospin between two particles
\begin{eqnarray}
P_\sigma &=& \frac{1}{2} (1+\vec{\sigma_1}\cdot\vec{\sigma_2}) ,~~~P_\tau = \frac{1}{2} (1+\vec{\tau_1}\cdot\vec{\tau_2}). \nonumber
\end{eqnarray}
For a zero-range interaction, 
\begin{equation}
 V(\mathbf{r_1},\mathbf{r_2})=V(|r_1-r_2|)\delta(\mathbf{r_1},\mathbf{r_2})
\end{equation}

the  matrix elements  in the two different channels $(S=0,T=1)$ and  $(S=1,T=0)$ read as 
\begin{equation}
\langle i\tau_i,\bar{i}\tau_{\bar{i}}|V^{T=1, S=0}| j\tau_j,\bar{j}\tau_{\bar{j}}\rangle=
\frac{1}{4}(\delta_{\tau_i \tau_j}\delta_{\tau_{\bar{i}} \tau_{\bar{j}}}+i\leftrightarrow\bar{i}) 
 \int d^3 r V ( r) \rho_i(r)\rho_j(r)
 \nonumber
\end{equation}

\begin{equation}
\langle i\tau_i,\bar{i}\tau_{\bar{i}}|V^{T=0, S=1}| j\tau_j,\bar{j}\tau_{\bar{j}}\rangle=
\frac{1}{4}(\delta_{\tau_i \tau_j}\delta_{\tau_{\bar{i}} \tau_{\bar{j}}}-i\leftrightarrow\bar{i}) 
 \int d^3 r V ( r) F_{i,j}(r)
 \nonumber
\end{equation}
where
\begin{equation}
 F_{i,j}(r)=2\Re \big[ \phi_i^*(+,r)  \phi_i(-,r)\phi_j(+,r)  \phi_j^*(-,r)\big]
 +\big[(\phi_i^*(+,r)  \phi_i(+,r)-\phi_i^*(-,r)  \phi_i(-,r)\big]
 \big[\phi_j(+,r)  \phi_j^*(+,r)  
 -\phi_j(-,r)  \phi_j^*(-,r)\big]. 
\end{equation}

\end{widetext}

\end{document}